\documentclass[amssymb,twocolumn,showpacs,aps]{revtex4b5}
\usepackage{epsfig}
\usepackage{amsmath}

\newcommand{\taueff}{\tau_{\text{eff}}}
\newcommand{\rhoeff}{\rho_{\text{eff}}}
\newcommand{\taumax}{\tau_{\text{max}}}

\def\Landau{\mathop{\cal O}\nolimits}

\begin{document}
 
\title{Hopping in the Glass Configuration Space: Subaging and
  Generalized Scaling Laws}

\author{Bernd Rinn,$^1$ Philipp Maass,$^{1,2}$ and Jean-Philippe Bouchaud$^2$}

\affiliation{$^1$Fachbereich Physik, Universit\"at Konstanz, 78457
  Konstanz, Germany} \affiliation{$^2$Service de Physique de l'Etat
  Condens\'e, CEA Saclay, 91191 Gif sur Yvette Cedex, France}

\date{April 17, 2001}

\begin{abstract}\vspace{0.3cm} 
  Aging dynamics in glassy systems is investigated by considering the
  hopping motion in a rugged energy landscape whose deep minima are
  characterized by an exponential density of states
  $\rho(E)=T_g^{-1}\exp(E/T_g)$, $-\infty\!<\!E\!\le\!0$. In
  particular we explore the behavior of a generic two-time correlation
  function $\Pi(t_w+t,t_w)$ below the glass transition temperature
  $T_g$ when both the observation time $t$ and the waiting time $t_w$
  become large. We show the occurrence of ordinary scaling behavior,
  $\Pi(t_w+t,t_w)\sim F_1(t/t_w^{\mu_1})$, where $\mu_1\!=\!1$ (normal
  aging) or $\mu_1\!<\!1$ (subaging), and the possible simultaneous
  occurrence of generalized scaling behavior,
  $t_w^{\gamma}[1-\Pi(t_w+t,t_w)]\sim F_2(t/t_w^{\mu_2})$ with
  $\mu_2\!<\!\mu_1$ (subaging). Which situation occurs depends on the
  form of the effective transition rates between the low lying states.
  Employing a ``Partial Equilibrium Concept'', the exponents
  $\mu_{1,2}$ and the asymptotic form of the scaling functions are
  obtained both by simple scaling arguments and by analytical
  calculations. The predicted scaling properties compare well with
  Monte--Carlo simulations in dimensions $d\!=\!1-1000$ and it is
  argued that a mean--field type treatment of the hopping motion fails
  to describe the aging dynamics in any dimension. Implications for
  more general situations involving different forms of transition
  rates and the occurrence of many scaling regimes in the $t$--$t_w$
  plane are pointed out.
\end{abstract}

\pacs{02.50.-r, 75.10.Nr, 05.20.-y}

\maketitle

\section{Introduction}\label{sec:introduction}

The history of glass formation strongly affects the relaxation
dynamics of glassy materials \cite{Zarzycki:1991,Parisi:1999}. This
dynamics is found to become slower with the ``age'' of the system,
that means with the time $t_w$ expired since the material was brought
into the glassy state. Such aging phenomena have been identified in
many systems and various dynamical probes. Prominent examples are
shear stress relaxations in structural glasses \cite{Struik:1978},
thermoremanent magnetizations or a.c.\ susceptibility in spin glasses
\cite{Vincent/etal:1997,Nordblad:1998}. Similar effects have been
observed on the dielectric constant of dipolar glasses
\cite{Alberici/etal:1997,Bouchaud/etal:2001}, of structural glasses
\cite{Leheny/Nagel:1998,Bellon/etal:2000}, and on the structure factor
of Lennard--Jones systems \cite{Barrat/Kob:2000}. More recent
experiments in colloidal gels \cite{Cipelliti/etal:2000} or other
`soft glassy materials' have been reported
\cite{Derec/etal:2000,Cloitre/etal:2001,Knaebel/etal:2000,Abou/etal:2001,Bellon/etal:2001},
and even electronic relaxations in Anderson insulators were found to
exhibit aging effects \cite{Vaknin/etal:2000}. Aging is also expected
for pinned systems (pinned domain walls
\cite{Vincent/etal:2000,Alberici/etal:1997}, pinned vortex lines
\cite{Nicodemi/Jenssen:2001}), polymer melts
\cite{Pitard/Shaknovitch:1999} and granular materials (see e.g.\ 
\cite{Barrat/Loreto:2001,Head:2000}).

From a theoretical point of view, several pictures have been proposed
\cite{Bouchaud/etal:1998}. The simplest one is based on domain
coarsening ideas \cite{Bray:1994}, and is probably well suited to
describe aging in, say, disordered ferromagnets where a well defined
order wants to establish across the system. However, in spin glasses
and even more evidently in glasses, the idea of some long range order
which progressively invades the system is far from trivial. Mean field
models for spin glasses, which are formally equivalent to the Mode
Coupling Theory of glasses \cite{Bouchaud/etal:1998}, do indeed lead
to aging phenomena below the glass transition. In this case, aging is
of geometric origin \cite{Laloux:1996}: As time grows, the system
progressively exhausts the possibilities of lowering its energy, and
finds itself around saddle points from which it is more and more
difficult to escape. Thermal activation is irrelevant in these models.
Although this picture might be justified for supercooled liquids {\it
  above} the Mode-Coupling temperature \cite{Cavagna:2001}, it
certainly breaks down at lower temperature, where activated events
become dominant. In this regime, one expects that a coarse-grained
dynamical model of thermally activated hops between metastable states
is a proper description of the dynamics. In fact, recent MD
simulations support this view
\cite{Donati/etal:2000,Angelani/etal:2000,Broderix/etal:2000}.
Landscape models have been widely discussed in the past
\cite{Goldstein:1969,Angell:1995}, but their relevance for aging
phenomena was recognized later
\cite{Bouchaud:1992,Feigelman/etal:1988,Bouchaud/Dean:1995,Monthus/Bouchaud:1996}.
These later developments were recently extended to treat rheological
phenomena \cite{Sollich/etal:1997}.

The `trap' models studied up to now lead to correlation or response functions
that depend on the ratio $t/t_w$ of the observation time $t$ to the waiting
time $t_w$ (full aging), or, for long-range correlated energy landscapes, on
the ratio $\log t/\log t_w$ \cite{Fisher:1999}.  However, many experimental
systems reveal {\it subaging} behavior, that is, the relevant variable is
$t/t_w^\mu$ with $\mu\!<\!1$.  Furthermore, it is possible that there exist,
for given waiting time $t_w$, various scaling regimes in time $t$, which are
governed by {\it different} relaxation times $\propto t_w^{\mu_s}$,
$s=1,2,\ldots$.  The occurrence of different scaling regions has recently been
conjectured on the basis of analytical results for mean-field spin glass
models \cite{Cugliandolo/Kurchan:1994,Bouchaud/etal:1998}. So far, however, it
was not possible to illustrate intuitively these multiple time regimes by
exact calculations on simpler models (see however \cite{Berthier/etal:2001}
for an interesting discussion of these multiple time scales). In this paper we
will discuss a model that allows us to demonstrate for the first time
explicitly the possible occurrence of subaging behavior and multiple time
scaling in a hopping model, where a point jumps among the deep (free) energy
minima $E_i$ of a complex configuration space. 
Some of the results discussed in this paper already appeared in a Letter
\cite{Rinn/etal:2000}.

\section{Hopping in a random energy landscape}\label{sec:definition}

\subsection{Model}
\label{sec:model}
\begin{figure}[t]
  \begin{center} \hspace*{-0.7cm}
    \epsfig{file=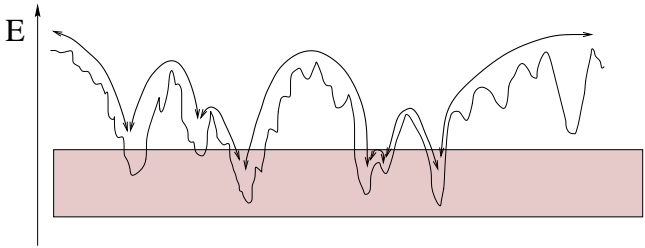,width=8.5cm} \caption{Sketch of
      a rugged energy landscape with various metastable minima. Within
      a coarse-grained description, the slow dynamics of the system
      may be attributed to effective ``super-transitions'' between the
      deepest minima belonging to the shaded area.}
    \label{fig:e-landscape}
    \end{center}
\end{figure}

Slow dynamics in glasses is often attributed to a thermally activated motion
of a point (henceforth denoted as ``particle'') that jumps among metastable
states in a rugged configuration space (see Fig.~\ref{fig:e-landscape}). In a
coarse grained description, only transitions between the deepest (free) energy
minima $E_i$ will govern the dynamics at long times. According to extreme
value statistics one may expect the distribution $\psi(E)$ of these deep
minima to be exponential (which is the behavior of the tail of a Gumbel
distribution, see e.g.\ \cite{Embrechts/etal:1997}). Indeed, mean-field
theories of spin glasses \cite{Bouchaud/Mezard:1997} and recent results from
molecular dynamics simulations \cite{Schoen/Sibani:1999} suggest this to be
the case.

For simplicity, we consider the metastable states with lowest energies
to be arranged on a hypercubic lattice in $d$ dimensions and refer to
them as ``sites''. The lattice here resembles an average finite
connectivity of the mutually accessible states. To each lattice site
$i$ is assigned a random energy $E_{i}$ drawn from a distribution
\begin{equation}
  \label{eq:Edistribution}
  \psi(E) = \frac{1}{T_{\text{g}}}\,
  \exp\left(\frac{E}{T_{\text{g}}} \right);\quad -\infty <
  E \le 0\ .
\end{equation}
As discussed in a moment, $T_{\text{g}}$ corresponds to a
``glass transition temperature'' and we thus define 
\begin{equation}
\theta\equiv T/T_{\text{g}} 
\label{eq:theta}
\end{equation}
as the rescaled temperature. Also, energies are specified in units of
$T_{\text{g}}$. The particle can jump from one site $i$ to any of the
$2d$ nearest neighbor sites $j$ with a hopping rate
\begin{equation}
  \label{eq:wij}
  w_{i,j} = \nu\, \exp\left(- \frac{[\alpha
  E_j-(1-\alpha)E_i]}{\theta}\right)\,, 
\end{equation}
where the ``attempt frequency'' $\nu\!\equiv\!1$ sets our time unit,
and the parameter $\alpha$ specifies how the energies of the initial
and target site contribute to the saddle point energy being surmounted
during a jump. In order for the $w_{i,j}$ to obey detailed balance,
$\alpha$ can assume any real value, but in a sense of a weighting of
the initial and target energy we restrict $\alpha$ to the range
\begin{equation}
0\le\alpha<1\,.
\label{eq:alpha}
\end{equation}
Note that the case $\alpha=0$ defines a trap model, where the jump
rates depend on the initial energy only, while the case $\alpha=1/2$
defines a ``force model'', where the jump rates are determined by the
energy difference between the two sites.

Due to the existence of very deep traps, the system exhibits a
``dynamical phase transition'' at the glass transition temperature
$\theta = 1$: in the high--temperature regime $\theta > 1$ the
Boltzmann distribution 
\begin{equation}
  \label{eq:normalization}
\psi(E)e^{-E/\theta}=e^{(1- 1/\theta) E}
\end{equation}
is normalizable, while in the low--temperature an equilibrium state
does not exist. In this latter situation the system is exploring
deeper and deeper traps as time proceeds and the overall dynamics
ages.

For the trap model ($\alpha = 0$) the dynamics is fully
characterized by the trapping times
\begin{equation}
\tau\equiv\exp(E/\theta)\,,
\label{eq:tau}
\end{equation}
with distribution
\begin{equation}
  \label{eq:trapping-times}
  \rho(\tau) = \theta\, \tau^{-1-\theta}\,,\hspace*{0.6cm}
1\le\tau<\infty
\end{equation}
The absence of an equilibrium state for $\theta < 1$ is reflected in
the fact that the mean trapping time $\langle \tau \rangle$ becomes
infinite. For $\alpha>0$ it is convenient to operate with the $\tau$
defined in (\ref{eq:tau}) as well, although these $\tau$ no longer have
the meaning of a trapping time.  The hopping rate (\ref{eq:wij}) can
then be written as
\begin{equation}
  \label{eq:wijtau}
  w_{i,j} = \frac{\tau_j^\alpha}{\tau_i^{1-\alpha}}\,.
\end{equation}

At first glance the model defined here seems to be similar to the
model considered earlier in \cite{Monthus/Bouchaud:1996}. However,
there is an important difference. In the model studied in
\cite{Monthus/Bouchaud:1996} the disorder is of ``annealed'' type,
i.~e.\ the energy of each site is drawn anew from $\psi(E)$ after each
jump. This can be viewed as a mean--field treatment and simplifies the
problem to a great extent. As will be shown in
sec.~\ref{sec:annealedcase}, in the annealed case one can map any
parameters $(\theta,\alpha>0)$ to parameters $(\theta',\alpha=0)$,
that means the $\alpha$ parameter is irrelevant. In the present model
by contrast the energy disorder is ``quenched'' and such a mapping is
not possible. We will show that this leads to the occurrence of much
richer aging dynamics, including subaging behavior and generalized
scaling forms.

\subsection{Aging function}
\label{subsec:agingfunction}

\begin{figure}[t]
  \begin{center}
    \hspace*{-0.7cm}
    \epsfig{file=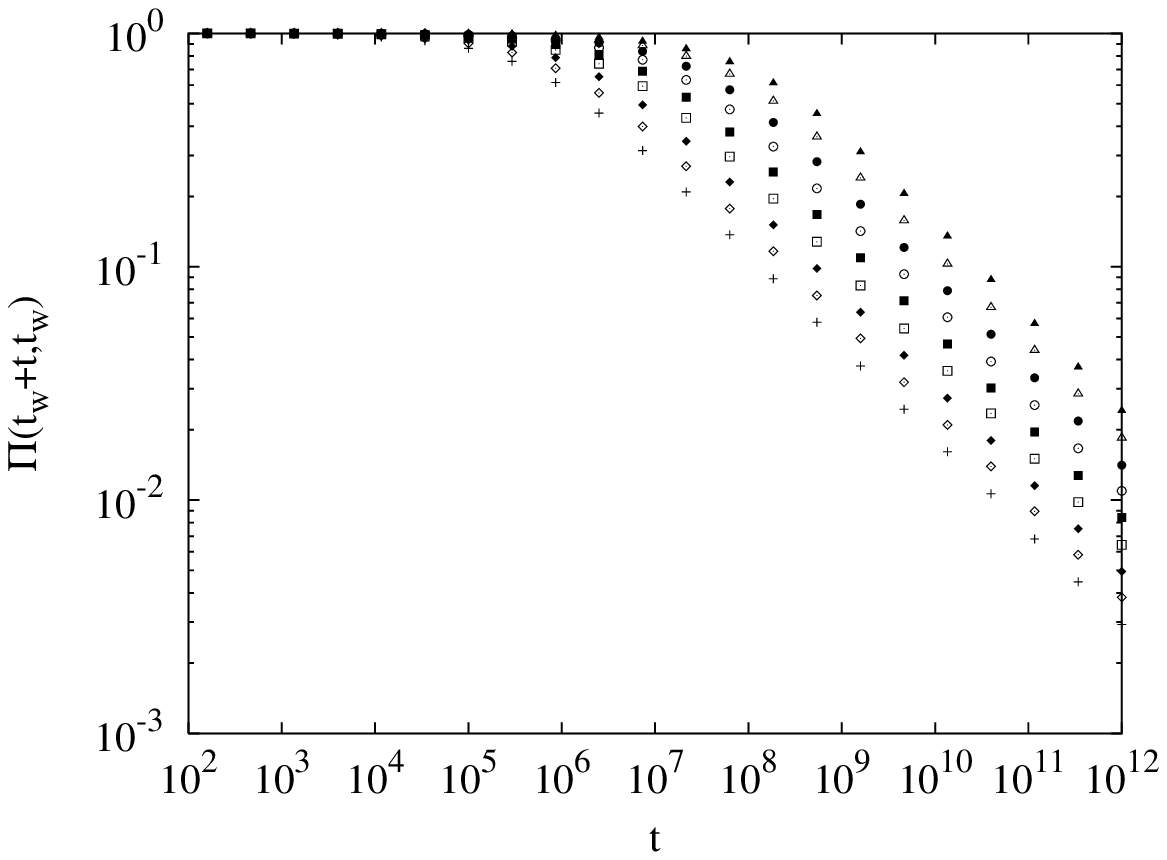,width=9.5cm}
    \caption{Aging function $\Pi(t+t_w,t_w)$ for $(d,\theta,\alpha) =
      (10,1/4,3/8)$ and different waiting times $t_w$. The symbols refer to
      waiting times $t_w=\, 2 \times 10^{8}\, ({\scriptstyle +})$, $5 \times
      10^{8}\, ({\scriptstyle \lozenge})$, $2 \times 10^{9}\, ({\scriptstyle
        \blacklozenge})$, $5 \times 10^{9}\, ({\scriptscriptstyle \square})$,
      $1 \times 10^{10}\, ({\scriptscriptstyle \blacksquare})$, $4 \times
      10^{10}\, (\circ)$, $1 \times 10^{11}\, (\bullet)$, $3 \times 10^{11}\,
      ({\scriptstyle \triangle})$, and $1 \times 10^{12}\, (\blacktriangle)$,
      respectively.}
    \label{fig:Afunscaled}
  \end{center}
\end{figure}
For studying aging properties we consider a quench from
$\theta=\infty$ to $\theta < 1$ at time 0, then wait for some time
$t_w\gg1$ and ask for the behavior of correlation functions followed
during an observation time $t$ after the waiting period. Since the
physical observables are functions of the coordinates of the
configuration space, they will essentially de-correlate after the
system has undergone a single transition from one deep minimum to
another one. Hence, a ``generic correlation function'' in the model is
to consider the probability that the particle has not jumped between
$t_w$ and $t_w+t$.  We denote this probability, after performing the
disorder average, by $\Pi(t_w+t,t_w)$. 

Let us note that $\Pi(t_w+t,t_w)$ can also be regarded
\cite{Bouchaud/Dean:1995} as the spin-spin-correlation function of the
generalized \textsc{Sherrington-Kirkpatrick} model with $p$-spin
interactions in the limit $p\to\infty$, which maps onto the random
energy model \cite{Derrida:1981}. It can also be viewed as an
incoherent intermediate scattering function at large wave numbers in
diffusion processes \cite{Monthus/Bouchaud:1996}.

A typical result for $\Pi(t+t_w,t_w)$ from continuous--time Monte
Carlo simulations is shown in Fig.~\ref{fig:Afunscaled} for parameters
$(d,\theta,\alpha) = (10, 1/4, 3/8)$ and different $t_w$ (for details
regarding the simulation see appendix \ref{sec:mc}). Indeed, we can
identify a pronounced aging phenomenon: The decay of $\Pi(t_w+t,t_w)$
with time $t$ becomes slower and slower with increasing waiting time
$t_w$.  The linear behavior of the curves at long times in the
double-logarithmic plot indicates a common asymptotic power--law decay,
and suggests that the various curves may be collapsed onto a common
master after rescaling the observation time $t$ by a proper function
of the waiting time $t_w$. The understanding of such scaling properties of
$\Pi(t+t_w,t_w)$ when both $t$ and $t_w$ become large ($t,t_w\gg1$)
will be the central issue in the following.

\section{Annealed situation}
\label{sec:annealedcase}

In the annealed case the model corresponds to a
continuous--time random walk (CTRW) irrespective of the value of
$\alpha$, since all times $\tau_j$ are drawn anew after each
transition. The whole process can be characterized by a
distribution $\rhoeff(\taueff)$ of effective trapping times
$\taueff$ (residence times between two transitions).

In order to calculate $\rhoeff(\taueff)$, let us consider the particle
to be at site $i=0$. Then
\begin{equation}
  \taueff = \frac{1}{\sum\limits_{j=1}^{2d}w_{0,j}}=
\frac{\tau_0^{1-\alpha}}{\sum\limits_{j=1}^{2d}\tau_j^\alpha}\,,
\label{eq:taueff}
\end{equation}
and the distribution is given by
\begin{eqnarray}
  \rhoeff(\taueff) &=& \left[ \prod_{i=0}^{2d} \int\limits_1^\infty d\tau_i\,
    \rho(\tau_i) \right]\, \delta\left( \taueff -
    \frac{\tau_0^{1-\alpha}}{\sum\limits_{j=1}^{2d} \tau_j^\alpha}
  \right) \nonumber\\
  &=& \left[ \prod_{i=0}^{2d} \int\limits_1^\infty d\tau_i\,
    \rho(\tau_i) \right]\, \frac{\taueff^{\alpha/(1-\alpha)}}{1-\alpha}
    \left[ \sum_{j=1}^{2d} \tau_j^{\alpha}
    \right]^{1/(1-\alpha)} \nonumber \\
    && \times\, \delta\left(\tau_0 -
      \taueff^{1/(1-\alpha)} \left[ \sum_{j=1}^{2d} \tau_j^{\alpha}
      \right]^{1/(1-\alpha)}  \right)\,.
\end{eqnarray}
The $\delta$ function contributes if
$\taueff\sum_{j=1}^{2d}\tau_j^\alpha\ge1$, i.e.\ there is no
restriction for $\taueff\ge1/2d$.  Hence, for $\taueff\ge 1/2d$:
\begin{equation}
  \rhoeff(\taueff) = C_{\text{eff}}\,
\theta'\,\taueff^{-1-\theta'}\,
\hspace{0.6cm}\theta'\equiv\frac{\theta}{(1-\alpha)}\,,
\label{eq:taueff-dist}
\end{equation}
where
$C_{\text{eff}}=\langle[\sum_{j=1}^{2d}\tau_j^\alpha]^{-\theta'}\rangle$,
and $\langle\ldots\rangle$ denotes an average over $2d$ uncorrelated
random numbers $\tau_j$ distributed according to
(\ref{eq:trapping-times}). The CTRW theory applies with a
rescaled temperature $\theta'$, and the aging properties of this model
have been worked out in \cite{Monthus/Bouchaud:1996}.

Note in particular that for $\theta'>1$ no aging occurs, even if the
distribution (\ref{eq:normalization}) is not normalizable for
$\theta<1$. This, however, is nothing to worry about, since in the
annealed model the energies change after each hop and the
distribution (\ref{eq:normalization}) does not correspond to an
equilibrium distribution. The dynamical phase transition in the
annealed model is defined by the diverging mean waiting time of the
distribution $\rhoeff(\taueff)$ that occurs for $\theta'<1$.  The
absence of aging for $0<1-\theta<\alpha$, that is when $\theta'>1$
but $0<\theta<1$, points to the fact that the annealed model may not
provide a valid mean field description of the quenched model in any
dimension $d$.

\begin{figure*}[t]
  \begin{center}
    \begin{minipage}{9.2cm}
      \hspace*{-0.4cm}
      \epsfig{file=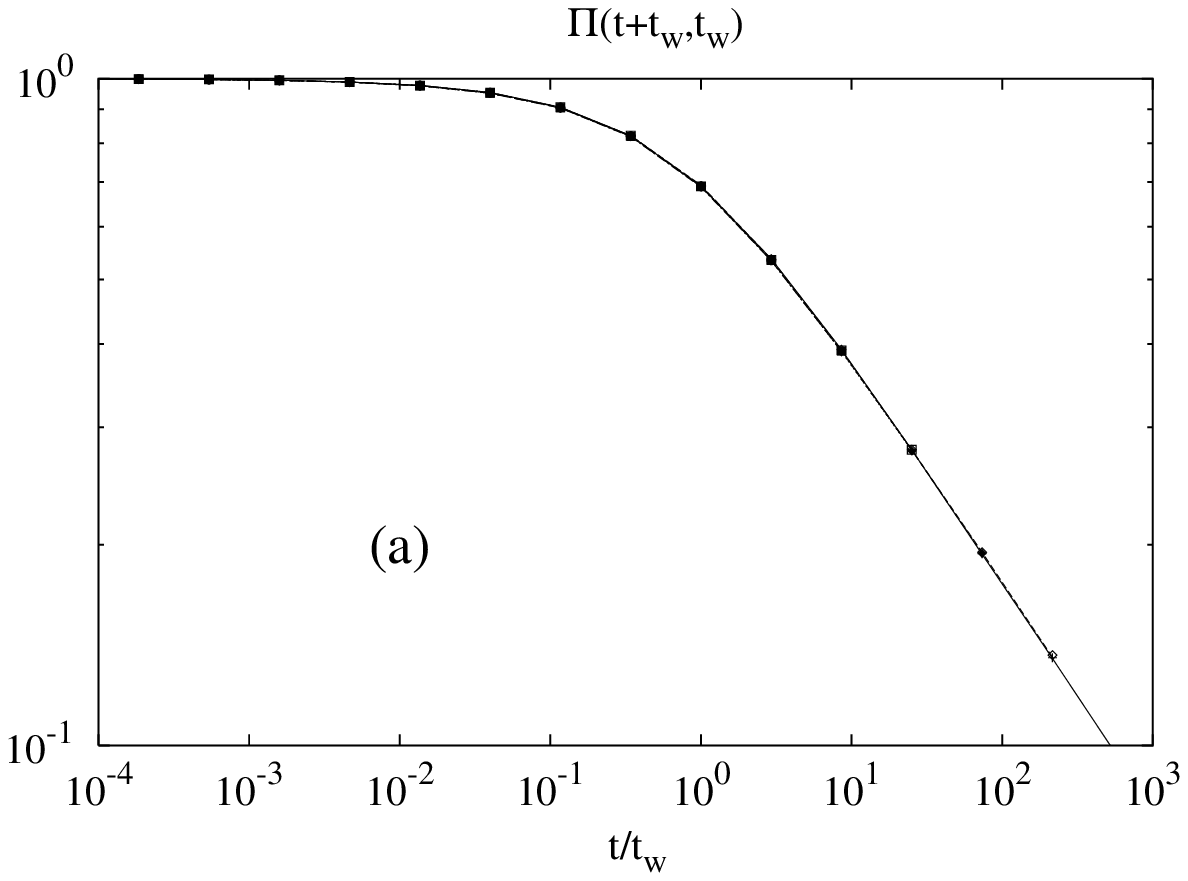,width=9.6cm}
    \end{minipage}
  \begin{minipage}{8.4cm}
    \epsfig{file=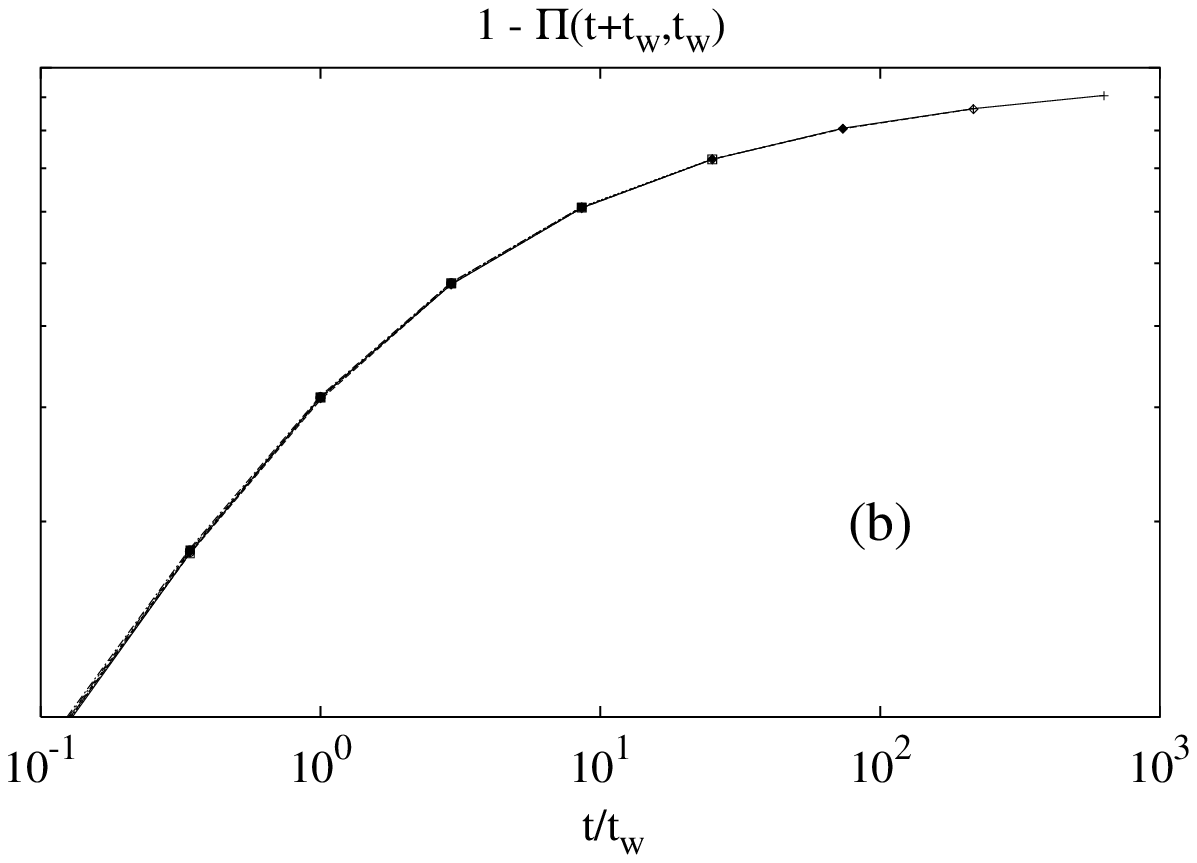,width=8.8cm,height=6.85cm}
  \end{minipage}
  \caption{Aging functions (a) $\Pi(t+t_w,t_w)$ 
    and its complement (b) $1-\Pi(t+t_w,t_w)$ from simulations of the
    annealed model with parameters $(d,\theta,\alpha) = (10,1/3,0)$
    (lines) and $(d,\theta,\alpha) = (10,1/4,1/4)$ (points). The
    waiting times range from $10^9$ to $10^{12}$. The scaling is so
    good that the various curves cannot be distinguished on the scale
    of the line thickness.}
  \label{fig:Afanscaled}
  \end{center}
\end{figure*}

In Fig.~\ref{fig:Afanscaled} the aging function $\Pi(t_w+t,t_w)$ is
shown for the annealed model as resulting from simulations for two
different parameters sets, $(d,\theta,\alpha) = (10,1/3,0)$ and
$(d,\theta,\alpha) = (10,1/4,1/4)$ that both refer to the same
$\theta'=1/3$. As expected, the aging functions for the two parameter
sets are indistinguishable. Moreover, as predicted in
\cite{Bouchaud/Dean:1995}, $\Pi(t_w+t,t_w)$ scales with $t/t_w$,
$\Pi(t_w+t,t_w)=F(t/t_w)$, where $F(u)\sim 1-u^{1-x}$ for $u\ll1$ (see
Fig.~\ref{fig:Afanscaled}b) and $F(u)\sim u^{-x}$ (see
Fig.~\ref{fig:Afanscaled}a). We will see in the following that this
simple behavior does not hold true any longer in the quenched case.
Nevertheless, since $\taueff$ is the inverse hopping rate on a site,
the formulae (\ref{eq:taueff},\ref{eq:taueff-dist}) will still be
useful in the following.

\section{Partial equilibrium concept (PEC)}\label{sec:pec}

Although full equilibrium cannot be reached in a system of infinite
size, there should be some equilibration on a local scale that
corresponds to the region of configuration space being explored by the
system after the quench. This is the idea of the PEC.

In the present model we can translate this idea into a precise though
approximate procedure for describing the aging process.  After the
waiting time $t_w$, the particle has visited $S=S(t_w)$ distinct sites
and we will assume that on these sites the system has equilibrated.
Accordingly, the probability $p_j$ to find the particle on the site
$j$ of the set of visited sites is
\begin{equation}
  p_j(S) = \frac{e^{-E_j/\theta}}{\sum_{k=1}^S e^{-E_k/\theta}}=
\frac{\tau_j}{\sum_{k=1}^S \tau_k}\ .
\end{equation}

The probability for the particle to remain on site $j$ for a time $t$ is
$\exp(-t\sum_{n_j} w_{j,n_j})=\exp(-t\,\tau_j^{\alpha-1}\sum_{n_j}
\tau_{n_j}^\alpha)$, where the sum over $n_j$ runs over all nearest neighbor
sites of $j$. Since the distinct visited sites form a Brownian path in the
lattice, which, on a local scale, has a one--dimensional topology, we will
consider exactly two of the $n_j$ sites to belong to the path. The remaining
$n-2$ sites are considered to be never visited until $t_w$. It will turn out
(see sec.~\ref{sec:exacteval}) that this assumption is not very important and
that the aging properties would be mainly the same if the Brownian path had a
compact structure (except for the asymptotics discussed in
sec.~\ref{subsec:l2limit}). According to the PEC, the aging function
$\Pi(t_w+t,t_w)$ is now approximated by taking the average of
$\exp(-t\,\tau_j^{\alpha-1}\sum_{n_j} \tau_{n_j}^\alpha)$ over all visited
sites $j$ with the weights $p_j$,
\begin{equation}
  \label{eq:pitilde}
  \tilde\Pi(t_w + t,t_w) \equiv \left\langle
    \frac{\sum_{j=1}^{S(t_w)}\tau_j
      \exp\Bigl(-t\,\tau_j^{\alpha-1}\sum_{n_j}
      \tau_{n_j}^\alpha\Bigr)}
    {\sum_{k=1}^{S(t_w)}\tau_k}
  \right\rangle\,.
\end{equation}
Here $\langle\ldots\rangle$ denotes an average over $(2d-1)\, S(t_w)$
uncorrelated random numbers $\tau_j$ that are distributed according to
(\ref{eq:trapping-times}). We will see that $\tilde\Pi(t_w + t,t_w)$
and $\Pi(t_w + t,t_w)$ exhibit the same scaling properties, which in
view of the generic character of $\Pi(t_w+t,t_w)$ (cf.\ 
Sec.~\ref{subsec:agingfunction}) are of primary interest for us here.
In quantitative terms both functions turn out to be different.  Since
$\tilde \Pi(t_w+t,t_w)$ depends on $t_w$ only through $S=S(t_w)$, we
will, in the following, also denote this function by $\tilde
\Pi(t,S)$.

As shown in appendix \ref{sec:properties}, the number of distinct
visited sites $S(t_w)$ grows with increasing waiting time $t_w$ as
\begin{subequations}
\begin{equation}
S(t_w)\sim t_w^\gamma\,,
\label{eq:stwgamma}
\end{equation}
where
\begin{equation}
\label{eq:gamma}
\gamma = \left\{
    {\renewcommand{\arraystretch}{1.5}
      \begin{array}{c@{\quad\text{for}\quad}l} 
        \displaystyle \frac{d\,\theta}{d\!+\!(2\!-\!d)\theta} & d < 2 \\
        \theta & d > 2 
      \end{array}} \right. \ .
\end{equation}
\end{subequations}
In $d=2$ there are logarithmic corrections, $S(t_w)\sim [t_w/\log
t_w]^\theta$.

\section{Simple scaling arguments}
\label{sec:scalingargument}

In this section we present simple arguments, why and how $\tilde
\Pi(t_w+t,t_w)$ is expected to scale.  In particular we propose
the presence of two characteristic
relaxation times $t_i(t_w)$, $i=1,2$, that grow as power laws with $t_w$,
\begin{equation}
t_i(t_w)\sim t_w^{\mu_i}\,,
\end{equation}
where $\mu_i$ are exponents depending on $\theta$ and $\alpha$.

\begin{figure}[t]
  \begin{center} \hspace*{-0.7cm}
    \epsfig{file=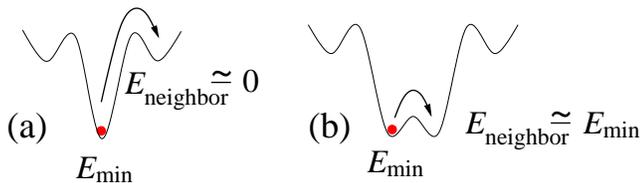,width=8.5cm} \caption{Jumps out
      of a valley with energy $E_{\rm min}$ after time $t_w$: (a) a
      situation corresponding to the typical situation, where the
      energies of the neighboring valleys have values close to zero;
      (b) a rare event where at least one of the neighboring valleys
      has an energy of order $E_{\rm min}$.}
    \label{fig:hop-illus}
\end{center}
\end{figure}

The first characteristic relaxation time $t_1(t_w)$ can be obtained as
follows.  After the waiting time $t_w$, the particle has visited $S=S(t_w)$
distinct sites and the occupation probability of these sites is dominated by
the one for the site with lowest energy $E_{\rm min}$ (in the glassy phase
$\theta<1$ the extreme values are dominant). The typical value of $E_{\rm
  min}$ can be estimated from $S\int_{-\infty}^{E_{\rm
    min}}\psi(E)=S\exp(E_{\rm min})\simeq1$, i.e.\ 
\begin{equation}
E_{\rm min}(t_w)\sim -\ln S(t_w)\sim -\gamma\ln (t_w)\,,
\end{equation}
or, in terms of the variable $\tau_{\rm max}$ corresponding to $E_{\rm
  min}$,
\begin{equation}
\tau_{\rm max}(t_w)=\exp(-E_{\rm min}/\theta)
\sim S(t_w)^{1/\theta}\sim t_w^{\gamma/\theta}\,.
\label{eq:taumax}
\end{equation}
Being at the site with energy $E_{\rm min}$, the particle typically
encounters a situation as drawn in Fig.~\ref{fig:hop-illus}a: All
neighboring sites have energies close to zero and the characteristic
escape time $t_1$ from the site with energy $E_{\rm min}$ is the
inverse hopping rate given in eq.~(\ref{eq:taueff}),
$t_1(t_w)\sim\tau_{\rm max}^{1-\alpha}\sim S^{(1-\alpha)/\theta}\sim
t_w^{\gamma(1-\alpha)/\theta}$, i.e.\
\begin{equation}
\mu_1=\frac{\gamma(1-\alpha)}{\theta}
\label{eq:mu1}
\end{equation}
Hence, for both $t$ and $t_w$ becoming large ($t,t_w\gg1$) and
$\Lambda_1=t/t_w^{\mu_1}$ being fixed
we expect $\tilde \Pi(t,S)$ to become a function of $\Lambda_1$ only,
\begin{equation}
\tilde \Pi(t, S) \sim F_1(\Lambda_1)\,,\hspace*{0.3cm}
\Lambda_1=t/t_w^{\mu_1}\,.
\label{eq:normalscal}
\end{equation}

The second relaxation time $t_2(t_w)$ is associated with rare events
depicted in Fig.~\ref{fig:hop-illus}b, which turn out to be important
at small times. The deepest state on an interval of length $S$ has a
typical value determined by the fact that it should occur with a
probability $1/S$. To obtain the scaling of the second deepest state
with $S$ we note that for a distribution exhibiting no peculiar
long--time tails, the gap between the deepest and second deepest state
remains finite when $S\to\infty$. Accordingly, with probability of
order $1/S\sim\tau_{\rm max}^{-\theta}$ one of the neighboring sites
can also have an energy comparable to $E_{\rm min}$.  In this case the
hopping rate specified in eq.~(\ref{eq:wijtau}) yields
$t_2\sim\tau_{\rm max}^{1-\alpha}/\tau_{\rm max}^\alpha\sim
S^{(1-2\alpha)/\theta}$, i.e.\ 
\begin{equation}
  \mu_2 = \frac{\gamma(1-2\alpha)}{\theta}\ .
\label{eq:mu2}
\end{equation}
Clearly this is a relevant time growing with $t_w$ only for $\alpha<1/2$.
Moreover, there is a further condition on the relevance of $t_2$. To see this,
we may consider the formal short--time expansion of $\tilde\Pi(t,S)$ from
eq.~(\ref{eq:pitilde}). Using the fact that a sum over random $\tau$'s raised
to some power with an exponent larger than $\theta$ scales as the maximal term
appearing in the sum (L\'evy statistics), we can estimate
\begin{eqnarray}
\label{eq:pi-shorttime}\phantom{.}
\hspace{-0.8cm}\tilde\Pi(t,S)&=&\left\langle\sum_{m=0}^{\infty}
\frac{(-t)^m}{m!}\sum_{j=1}^S p_j\, \tau_j^{-(1-\alpha)m}
\Bigl(\sum_{n_{j}}\tau_{n_j}^\alpha\Bigr)^m\right\rangle\nonumber\\
&\simeq&1+\sum_{m=1}^{\infty}
\frac{(-t)^m}{m!\,\tau_{\rm max}^{(1-\alpha)m}}\,
\Bigl[1+\tau_{\rm max}^{-\theta}\tau_{\rm max}^{\alpha m}\Bigr]\,.
\end{eqnarray}
Here the first term in the rectangular brackets corresponds to the typical
situation, while the second term corresponds to the rare events and is
weighted by a factor $\tau_{\rm max}^{-\theta}$ (we have neglected constant
prefactors). Comparing these two terms, we see that, if $\theta<\alpha$, the
latter term dominates for {\it all} $m$.  In this case we thus expect the
scaling behavior of the form
\begin{equation}
    S\, \big[ 1-\tilde \Pi(t,S) \big] \sim F_2(\Lambda_2)\,,
\hspace{0.4cm}\Lambda_2=t/t_2(t_w)
\label{eq:genscal}
\end{equation}
for times $t,t_w\gg1$ and $\theta<\alpha<1/2$.

For $\alpha<\theta$ to the contrary, the first term in the rectangular
brackets dominates for small $1\le m\le\theta/\alpha$, and $t_2(t_w)$ is
not significant [note that $t_2(t_w)$ is smaller than $t_1(t_w)$]. The fact
that eq.~(\ref{eq:pi-shorttime}) is not a regular expansion in the scaling
variable $t/t_1(t_w)=t/\tau_{\rm max}^{1-\alpha}$ (for $\alpha>0$) indicates
that $F_1(\Lambda_1)$ does not have an analytical behavior for small
$\Lambda_1$. On the other hand, the formal short--time expansion suggests a
regular behavior $F_2(\Lambda_2)\sim \Lambda_2$.  An exact treatment of the
PEC formula (\ref{eq:pitilde}) presented in sec.~\ref{sec:exacteval}, however,
shows that this should be true only in $d=1$. The small $\Lambda_2$ regime
is in fact a subtle one since it is very sensitive to rare events and
the connectivity properties of the Brownian path.

We can furthermore deduce the asymptotic form of $F_1(\Lambda_1)$ for
$\Lambda_1\to\infty$ and $\Lambda_1\to0$ by simple arguments as will be shown
next. The behavior of $F_2(\Lambda_2)$ for large $\Lambda_2$ then follows from
the fact that the large $\Lambda_2$ behavior of eq.~(\ref{eq:genscal}) should
match the small $\Lambda_1$ behavior of eq.~(\ref{eq:normalscal}), i.\,e.\ 
$1-S^{-1}F_2(t/t_w^{\mu_2})\simeq F_1(t/t_w^{\mu_1})$ for $1\ll t_w^{\mu_2}\ll
t\ll t_w^{\mu_1}$.

\subsection{Limit $\Lambda_1 \to \infty$}

The particle leaves the site reached after $t_w$ typically in a time
$t_w^{\mu_1}$. In order to explore the behavior for large
$\Lambda_1=t/t_w^{\mu_1}\gg1$ we may assume $t\gg t_w\gg1$ and ask,
which events give rise to a non--vanishing $\tilde\Pi(t,S)$. These are
\emph{rare} events, where an unusual large inverse hopping rate
$\taueff \geq t+t_w\simeq t$ has been encountered before the time
$t_w$ has passed, or, said differently, before the particle has
visited $S(t_w)$ distinct sites.  The probability that $\taueff$ on
one site is smaller than $t$ is ${\cal P}(t)=\int_0^t
d\taueff\,\rhoeff(\taueff)\sim 1-t^{-\theta/(1-\alpha)}$ and the
probability that at least for one of $S=S(t_w)$ sites $\taueff$ is
larger than $t$ is $1-{\cal P}^S$. Hence,
\begin{equation}
  \tilde\Pi(t,S) \sim 1 - {\cal P}^S = 1 - (1- {\cal Q})^S \sim S(t_w)\,
{\cal Q}(t)
\end{equation}
where ${\cal Q}(t)=1-{\cal P}(t)\sim t^{-\theta/(1-\alpha)}$.
Inserting $S(t_w)$ from eq.~(\ref{eq:stwgamma})
we thus obtain
\begin{equation}
  F_1(\Lambda_1) \sim \Lambda_1^{-\delta}\,,\hspace{0.3cm}
  \delta \equiv \frac{\theta}{1-\alpha}\ .
\label{eq:f1lamtoinf}
\end{equation}

\subsection{Limit $\Lambda_1 \to 0$}

In the limit $\Lambda_1\to 0$ corresponding to $1\ll t\ll t_w^{\mu_1}$, the
\emph{typical} situations govern the decay of $\tilde\Pi(t_w+t,t_w)$.  In
these typical situations, energies $E\ll E_{\rm min}(t_w)$ or Boltzmann
factors $\tau\gg\tau_{\rm max}(t_w)$ do not matter.  Let us then consider a
jump of the particle from a site with energy $E_1$ [Boltzmann factor
$\tau_1=\exp(-E_1/\theta)$] to a neighboring site with energy $E_2$ [Boltzmann
factor $\tau_2=\exp(-E_2/\theta)$], where $0\le E_1,E_2\le E_{\rm min}$.  On
average such a jump occurs at a time $\tau_2^\alpha/\tau_1^{1-\alpha}$.  The
probability for the particle to be at a site with energy in an interval
$(E_1,E_1+dE_1)$ is
$\psi(E_1)\exp(-E_1/\theta)dE_1\propto\tau_1^{-1-\theta}\tau_1 d\tau_1$
(``equilibrated initial site'') and the probability for the energy of a
neighboring site to be in an interval $(E_2,E_2+dE_2)$ is
$\psi(E_2)dE_2\propto\tau_2^{-1-\theta}d\tau_2$ (``random target site'').
Hence the probability $\psi(t)dt$ for the particle to leave the site reached
after $t_w$ in a time interval $(t,t+dt)$ can be estimated by
\begin{eqnarray}
  \psi(t,\taumax) &\propto& \int_1^{\taumax}\hspace*{-0.4cm} d\tau_1\,
  \tau_1^{-\theta} \int_1^{\taumax} \hspace*{-0.4cm}d\tau_2\,
\tau_2^{-1-\theta}
\delta\left(t -
    \frac{\tau_1^{1-\alpha}}{\tau_2^\alpha}\right)\nonumber\\[0.15cm]
&=:& \phi(t,\taumax)\ .
\label{eq:phit} 
\end{eqnarray}
Normalizing $\psi(t)$ on its support $\taumax^{-\alpha} \le t \le
\taumax^{1-\alpha}$ we obtain
\begin{equation}
  \label{eq:psieff}
  \psi(t,\taumax) = \frac{\phi(t,
  \taumax)}{\int\limits_{\taumax^{-\alpha}}^{\taumax^{1-\alpha}} 
  dt'\, \phi(t', \taumax)} \ .
\end{equation}
Since $1-\tilde\Pi(t_w+t,t_w)$ is the probability of the particle to
leave the site reached after $t_w$ within the time interval 
$[0,t)$, we expect
\begin{equation}
  \label{eq:onempi}
  1-\tilde \Pi(t_w+t,t_w)\sim \int_0^t\, dt'\,
  \psi\big(t',\taumax(t_w)\big)
\end{equation}
for $t,t_w\gg1$ and $t/t_w^{\mu_1}\ll1$. Performing the integrals
in eq.~(\ref{eq:phit}) gives
\begin{equation}
  \label{eq:phi}
  \phi(t,\taumax) = \frac{1-\alpha}{\alpha-\theta}\,
  {t}^{\frac{\alpha-\theta}{1-\alpha}} \left[ \left(
      \frac{t}{\taumax^{1-\alpha}}
    \right)^{-\frac{\alpha-\theta}{(1-\alpha)\,\alpha}}  - 1 \right]\ . 
\end{equation}
Since $t/\taumax^{1-\alpha}\sim\Lambda_1$, we have, in the limit
$\Lambda_1\to 0$, to distinguish between $\alpha > \theta$ and
$\alpha < \theta$, yielding
\begin{equation}
  \label{eq:phi2}
  \phi(t, \taumax)\sim \left\{
    {\renewcommand{\arraystretch}{2.5}
    \begin{array}{l@{\quad}l}
      \displaystyle \frac{1-\alpha}{\alpha-\theta}\, {t}^{-1 +
        \theta/\alpha}\, \taumax^{1-\theta/\alpha}\,, & \alpha > \theta \\
      \displaystyle \frac{1-\alpha}{\theta-\alpha}\, {t}^{-1 +
        \frac{1-\theta}{1-\alpha}}\,, & \alpha < \theta
    \end{array}} \right. \ .
\end{equation}
Inserting this into eqs.~(\ref{eq:psieff} \ref{eq:onempi}) and using
(\ref{eq:taumax}) then gives
\begin{subequations}
\label{eq:f1lamtozero}
  \begin{equation}
    1 - F_1(\Lambda_1) \sim \Lambda_1^{\varepsilon}\,,\quad\quad
\Lambda_1\to0
\label{eq:f1lamtozeroa}
  \end{equation}
  with
  \begin{equation}
    \varepsilon=\left\{ 
    {\renewcommand{\arraystretch}{2.0}
      \begin{array}{l@{\quad}l}
        \displaystyle \varepsilon_>\, \equiv\, 
        \frac{\theta}{\alpha}\,, &
        \alpha > \theta\\[0.1cm] 
        \displaystyle \varepsilon_<\, \equiv\,
        \frac{1-\theta}{1-\alpha}\,, & \alpha < \theta
      \end{array}} \right. \ .
    \label{eq:f1ltozerob}
  \end{equation}
\end{subequations}

As discussed above, by matching this small $\Lambda_1$ behavior of
$F_1(\Lambda_1)$ to the large $\Lambda_2$ behavior of $F_2(\Lambda_2)$ we
furthermore obtain for $\alpha>\theta$
\begin{equation}
  F_2(\Lambda_2) \sim \Lambda_2^{\theta/\alpha}\,,
  \quad\quad \Lambda_2 \to \infty\ .
  \label{eq:f2lamtoinf}
\end{equation}

\section{Test of the PEC}

\begin{figure*}[t]
  \begin{center}
    \begin{minipage}{9.2cm}
      \hspace*{-0.4cm}
      \epsfig{file=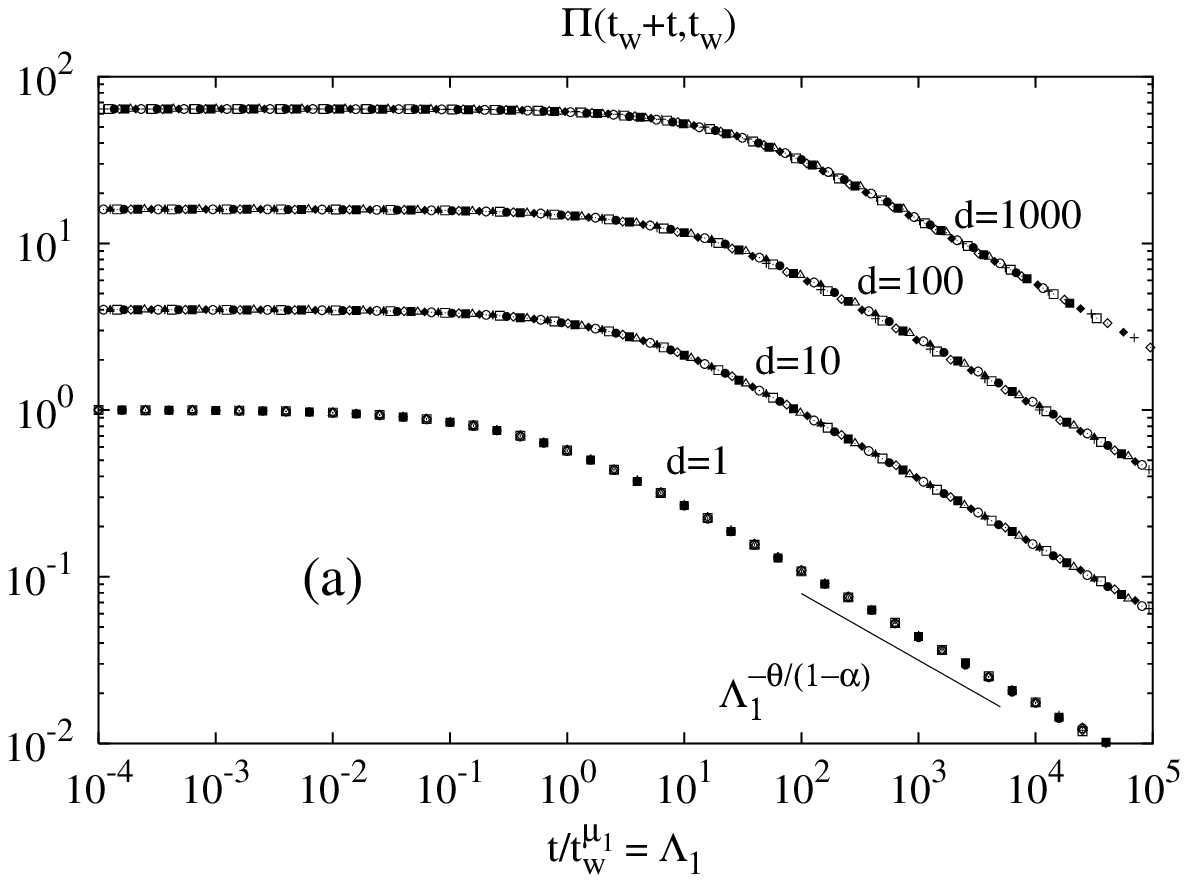,width=9.6cm}
    \end{minipage}
  \begin{minipage}{8.4cm}
    \epsfig{file=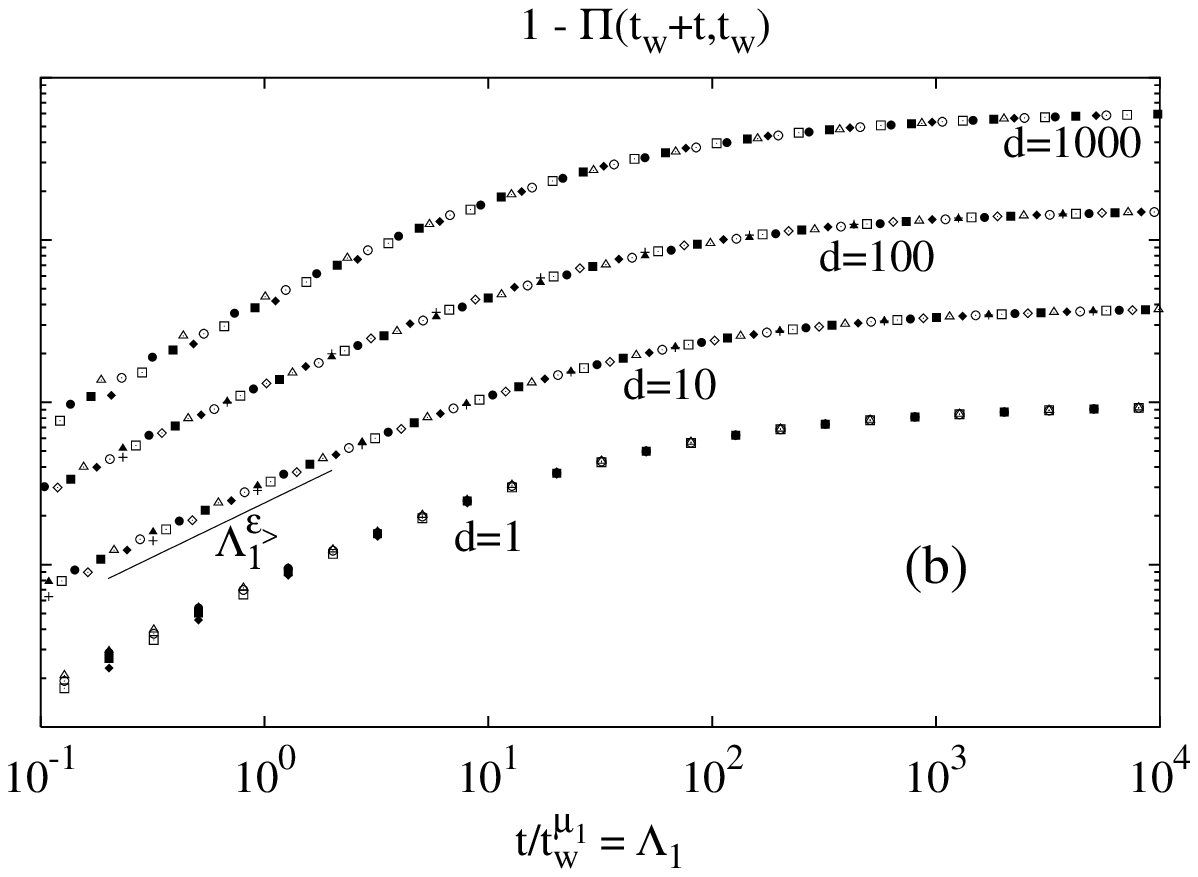,width=8.8cm,height=6.85cm}
  \end{minipage}
  \caption{First scaling function for
    a parameter set with $\alpha>\theta$: (a) $\Pi(t_w+t,t_w)$ and (b)
    $1-\Pi(t_w+t,t_w)$ as a function of $t/t_w^{\mu_1}=\Lambda_1$ for
    parameters $(\theta,\alpha) = (1/4,3/8)$ and different dimensions
    $d$. The straight lines indicate the asymptotic behavior according
    to eqs.~(\ref{eq:f1lamtoinf},\ref{eq:f1lamtozero}).  In order to
    make the graphs distinguishable, $\Pi(t_w+t,t_w)$ and
    $1-\Pi(t_w+t,t_w)$ have been multiplied by factors 4, 16
    and 64 for $d=10,100$ and 1000, respectively. In plot (b) also
    $t/t_w^{\mu_1}$ has been multiplied by factors 32, 4, 1 and
    0.5 for $d=1$, 10, 100 and 1000, respectively. For $d=10$ and
    $d=100$ the symbols refer to the same waiting times as in
    Fig.~\ref{fig:Afunscaled}. For $d=1$ the symbols refer to $t_w=\,
    6 \times 10^{6}\, ({\scriptstyle +})$, $2 \times 10^{7}\,
    ({\scriptstyle \lozenge})$, $4 \times 10^{7}\, ({\scriptstyle
      \blacklozenge})$, $1 \times 10^{8}\, ({\scriptscriptstyle
      \square})$, $3 \times 10^{8}\, ({\scriptscriptstyle
      \blacksquare})$, $6 \times 10^{8}\, (\circ)$, $2 \times 10^{9}\,
    (\bullet)$, $4 \times 10^{9}\, ({\scriptstyle \triangle})$, and $1
    \times 10^{10}\, (\blacktriangle)$. For $d=1000$ the symbols refer
    to $t_w=\, 6 \times 10^{6}\, ({\scriptstyle \blacklozenge})$, $1
    \times 10^{7}\, ({\scriptscriptstyle \square})$, $3 \times
    10^{7}\, ({\scriptscriptstyle \blacksquare})$, $8 \times 10^{7}\,
    (\circ)$, $2 \times 10^{8}\, (\bullet)$, and $4 \times 10^{8}\,
    ({\scriptstyle \triangle})$.}
  \label{fig:Afscaleda}
  \end{center}
\end{figure*}

\begin{figure*}[t]
  \begin{center}
    \begin{minipage}{9.2cm}
      \hspace*{-0.4cm}
      \epsfig{file=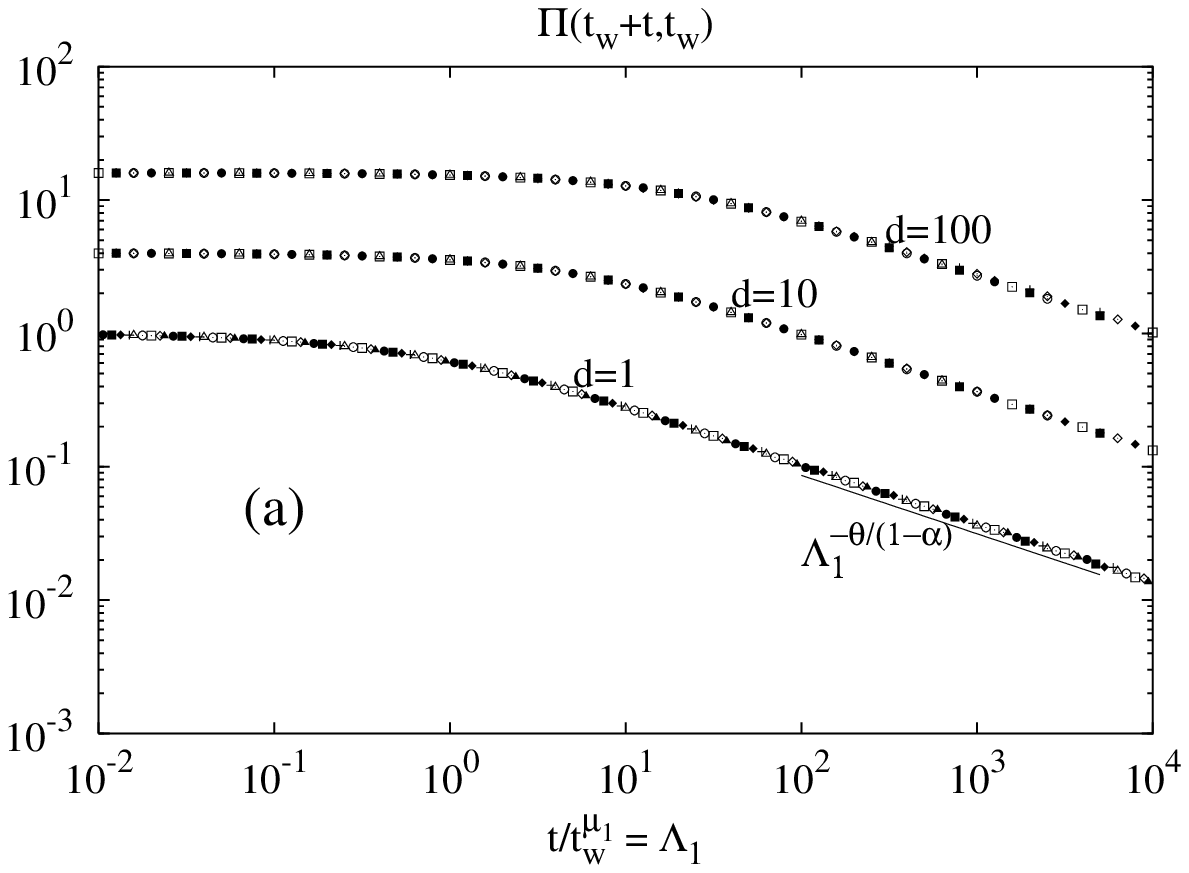,width=9.6cm}
    \end{minipage}
    \begin{minipage}{8.4cm}
      \epsfig{file=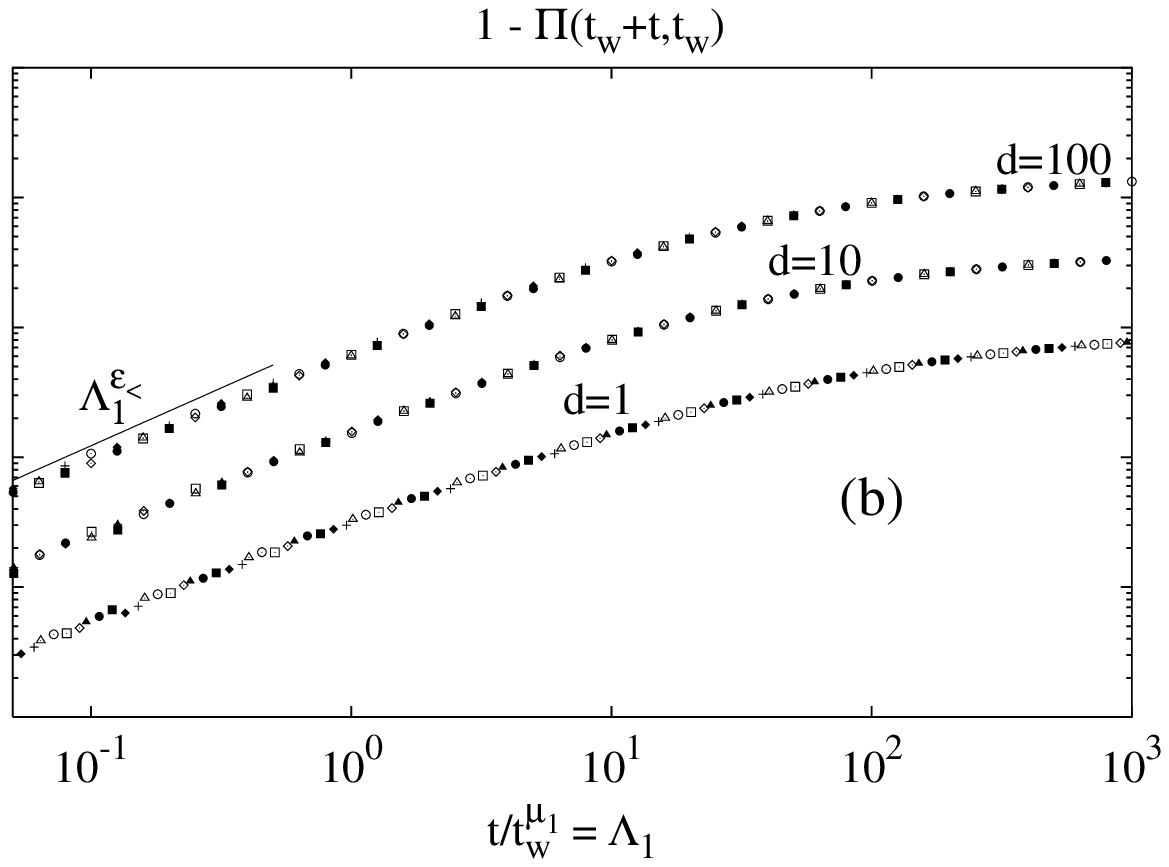,width=8.8cm,height=6.85cm}
    \end{minipage}
    \caption{First scaling function for
      a parameter set with $\alpha<\theta$: (a) $\Pi(t_w+t,t_w)$ and
      (b) $1-\Pi(t_w+t,t_w)$ as a function of
      $t/t_w^{\mu_1}=\Lambda_1$ for $(\theta,\alpha)=(1/3,1/4)$ and
      different dimensions $d$.  The straight lines indicate the
      asymptotic behavior according to
      eqs.~(\ref{eq:f1lamtoinf},\ref{eq:f1lamtozero}). In order to
      make the graphs distinguishable $\Pi(t_w+t,t_w)$ and
      $1-\Pi(t_w+t,t_w)$ have been multiplied by factors 4 and
      16 for $d=10$ and 100, respectively. In plot (b) also
      $t/t_w^{\mu_1}$ has been multiplied by factors 32 and 4 for
      $d=1$ and 10, respectively. The symbols refer to the same
      waiting time as for $d=1$ in Fig.~\ref{fig:Afscaleda}.}
    \label{fig:Afscaledb}
  \end{center}
\end{figure*}

\begin{figure}[b]
  \begin{center}
    \hspace*{-0.7cm}
    \epsfig{file=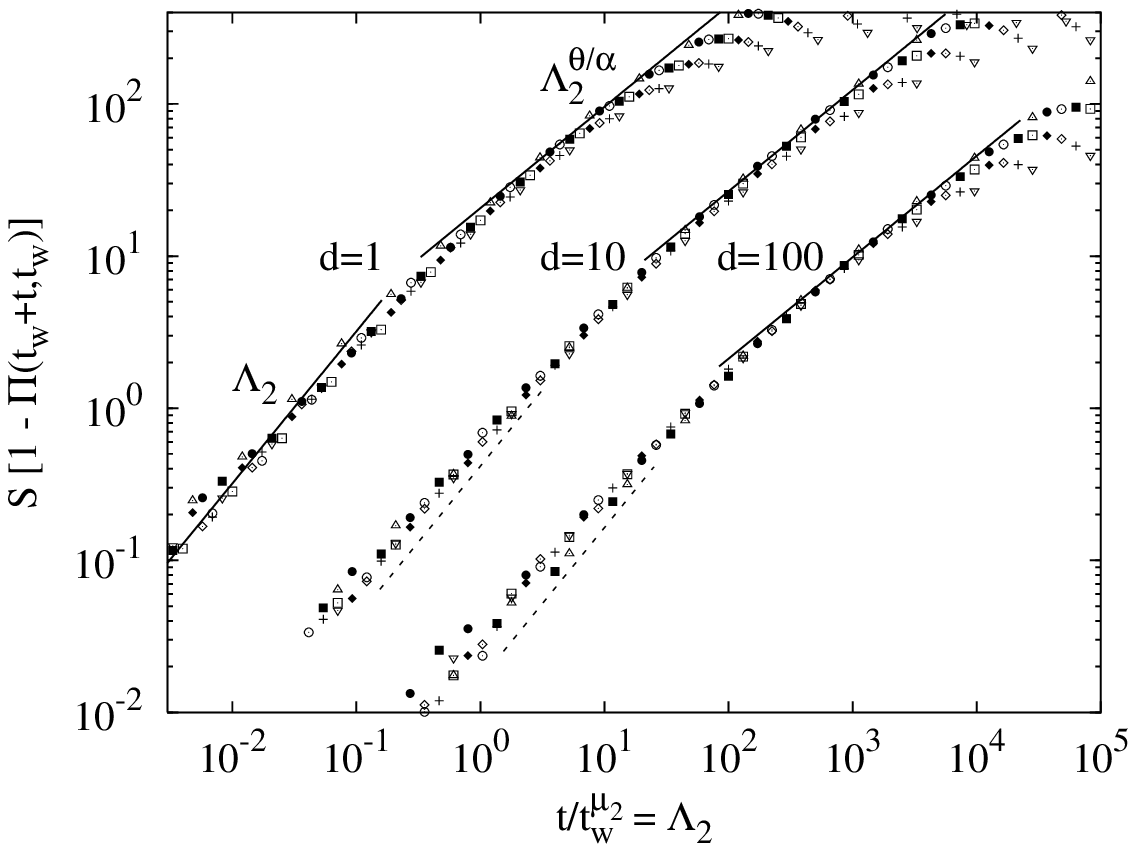,width=9.5cm}
    \caption{Scaling function $F_2(\Lambda_2)$ for the same parameter
      set as in Fig.~\ref{fig:Afscaledb}: $S[1-\Pi(t_w+t,t_w)]$ as a function
      of $t/t_w^{\mu_2}=\Lambda_2$ for $(\theta,\alpha)=(1/4,3/8)$ and $d=1$,
      10, 100.  The straight lines for large $\Lambda_2$ indicate the
      asymptotic behavior according to eq.~(\ref{eq:f2lamtoinf}), while the
      solid ($d=1$) and dashed lines ($d=10,100$) for small $\Lambda_2$
      correspond to a linear behavior (see text).  The symbols refer to the
      same waiting times as in Fig.~\ref{fig:Afscaleda}, with the additional
      waiting time $t_w=3 \times 10^6$. In order to make the graphs
      distinguishable, $S[1-\Pi(t_w+t,t_w)]$ has been multiplied by factors 10
      and 5 for $d=1$ and 10, respectively.}
    \label{fig:Afscaled2}
  \end{center}
\end{figure}

\subsection{Scaling properties}

In order to test the PEC, we performed simulations of the model in
$d=1,10,100$ and $1000$ by means of a continuous--time Monte--Carlo
algorithm (see appendix \ref{sec:mc} for details of the simulation
procedure). Averages were taken over $10^5$ (for $d=1,10,100$) and
$10^4$ (for $d=1000$) energy landscapes.

Figures \ref{fig:Afscaleda}a and \ref{fig:Afscaleda}b show,
respectively, $\Pi(t+t_w,t_w)$ and $1-\Pi(t+t_w,t_w)$ as a function of
the scaled variable $\Lambda_1=t/t_w^{\mu_1}$ (cf.\ 
eqs.~(\ref{eq:mu1},\ref{eq:normalscal}) for $(\theta,\alpha) =
(1/4,3/8)$, i.e.\ a case where $\alpha>\theta$. As expected from the
PEC and the scaling arguments outlined in the previous section, the
data collapse onto master curves $F_1(\Lambda_1)$ for all dimensions
$d$.  In particular we find $F_1(\Lambda_1)\sim\Lambda_1^{\epsilon_<}$
for $\Lambda_1\to0$ and $F_1(\Lambda_1)\sim\Lambda_1^{-\delta}$ in
agreement with eqs.~(\ref{eq:f1lamtozero}a,b) and (\ref{eq:f1lamtoinf}),
respectively.

Correspondingly scaled data for $(\theta,\alpha) = (1/3,1/4)$, i.e.\ a
case where $\alpha<\theta$, are shown in fig.~\ref{fig:Afscaledb}.
Again there is a good data collapse and the scaling functions
$F_1(\Lambda_1)$ have the expected asymptotic behavior for small and
large $\Lambda_1$. In particular we now find
$F_1(\Lambda_1)\sim\Lambda_1^{\epsilon_>}$ with $\epsilon_>$ from
eq.~(\ref{eq:f1lamtozeroa}).

Moreover, for $\alpha>\theta$, PEC and the scaling arguments predict the
presence of a second time scale $t_2=t_w^{\mu_2}$ [cf.\ eq.~(\ref{eq:mu2})]
and an associated generalized scaling $t_{w}^{\gamma}[1-\Pi(t+t_w,t_w)]\sim
F_2(t/t_w^{\mu_2})$ [cf.\ eqs.~(\ref{eq:genscal},\ref{eq:stwgamma})].  The
occurrence of this generalized scaling is verified in fig.~\ref{fig:Afscaled2}
for $d=1$, 10 and 100. The master curves scale
$F_2(\Lambda_2)\sim\Lambda_2^{\theta/\alpha}$ for $\Lambda_2 \to \infty$ as
predicted by eq.~(\ref{eq:f2lamtoinf}). The critical reader may note that the
simulated data do not collapse at large $\Lambda_2$, the deviations from
scaling setting in at larger $\Lambda_2$ for larger $t_w$. The reason for this
is that the limits $t_w\to \infty$ and $\Lambda_2=t/t_w^{\mu_2}\to\infty$ must
not be commuted. One first has to take the limit $t_w\to\infty$ to obtain the
scaling function $F_2(\Lambda_2)$ and then has to consider the asymptotic
behavior for large $\Lambda_2$.

The behavior for $\Lambda_2\to0$ is not so clear. For $d>1$, the exact
treatment of the PEC formula (\ref{eq:pitilde}) yields
$F_2(\Lambda_2)\sim\Lambda_2^{\theta/\alpha}$ also (with a smaller
prefactor in the scaling law), while for $d=1$,
$F_2(\Lambda_2)\sim\Lambda_2$ seems to be correct. In fact, for $d=1$
the data in fig.~\ref{fig:Afscaled2} are in fair agreement with a
linear behavior. For $d=10,100$ the data also show some linear
dependence for intermediate $\Lambda_2$ values (see the dashed lines
in fig.~\ref{fig:Afscaled2}), and in fact, such intermediate regime is
predicted to occur based on the PEC formula (although it should be
less pronounced, see the discussion in sec.~\ref{subsec:l2limit}). For
very small $\Lambda_2$, however, we expect
$F_2(\Lambda_2)\sim\Lambda_2^{\theta/\alpha}$, although this
asymptotic regime can not be identified in fig.~\ref{fig:Afscaled2}.
At best we can say that there is a change in curvature at very small
$\Lambda_2$. The true asymptotics, however, could not be obtained
within reasonable CPU time.

Accounting for the existence of two different time scales is very important to
properly rescale the numerical results. Had one assumed a single time scale
$t_w^\mu$, one would have obtained an approximate data collapse with an
effective value of $\mu$ intermediate between $\mu_1$ and $\mu_2$. This remark
might be of importance for analyzing experimental data: the assumption of a
single time scale could lead to a systematic underestimation of the true,
asymptotic value of $\mu$ (see the discussion in \cite{Bouchaud:2000}).

\subsection{Full comparison with simulations}

\begin{figure*}[t]
  \begin{center}
    \begin{minipage}{9.2cm}
      \hspace*{-0.4cm}
      \epsfig{file=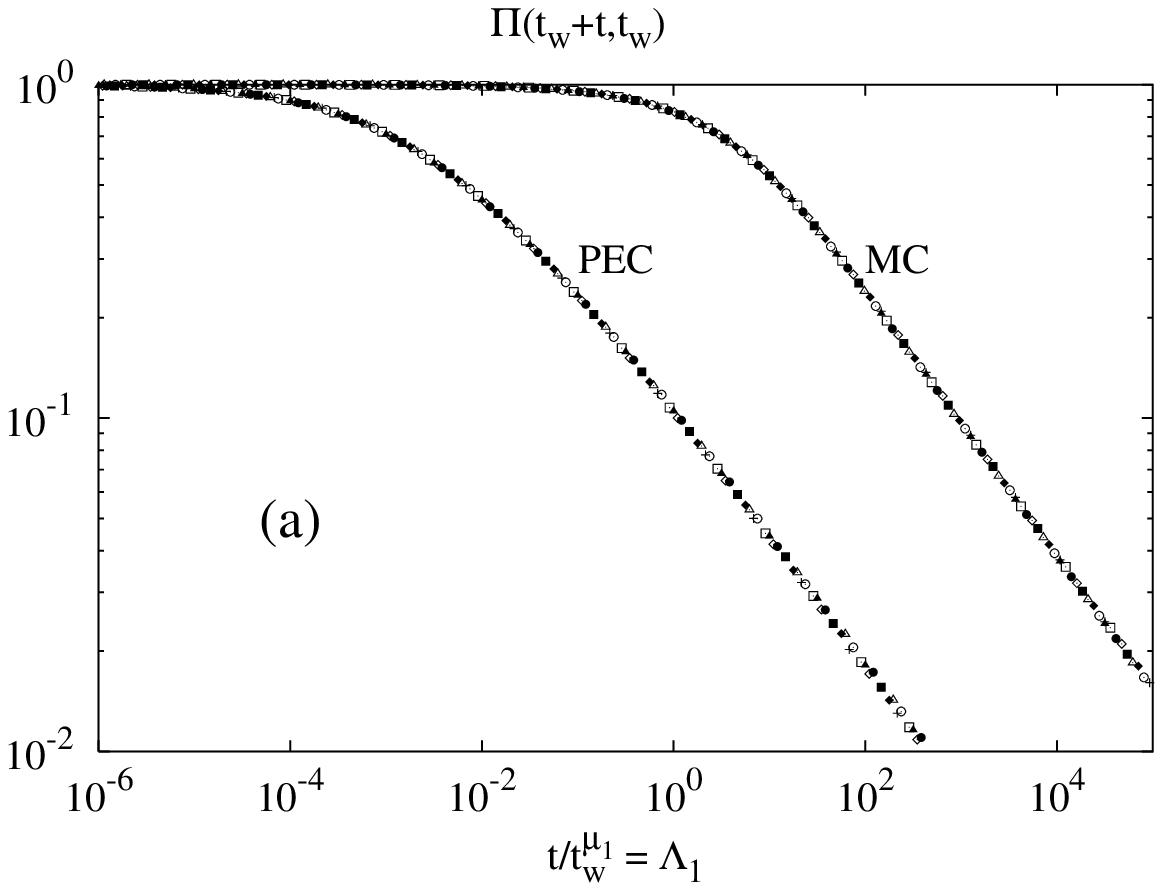,width=9.6cm}
    \end{minipage}
    \begin{minipage}{8.4cm}
      \epsfig{file=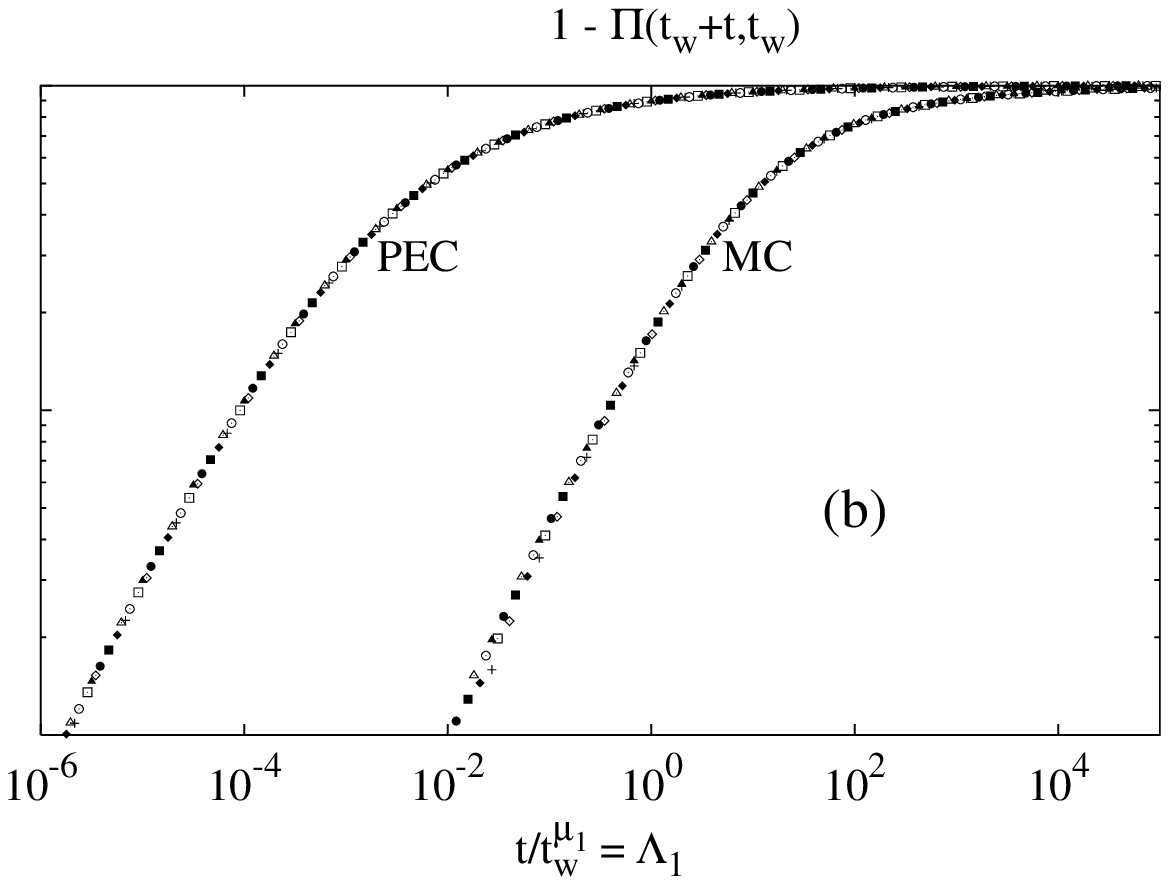,width=8.9cm,height=6.85cm}
    \end{minipage}
    \caption{$\Pi(t_w+t,t_w)$ and $1-\Pi(t_w+t,t_w)$ for
      $(d,\theta,\alpha)=(10,1/4,3/8)$ as a function of the scaling
      variable $t/t_w^{\mu_1}$ from Monte--Carlo simulations (MC) and
      the Partial Equilibrium formula, eq.~(\ref{eq:pitilde}) (PEC).
      The symbols refer to waiting times $t_w=\, 2 \times 10^{11}\,
      ({\scriptstyle +})$, $6 \times 10^{11}\, ({\scriptstyle
        \lozenge})$, $3 \times 10^{12}\, ({\scriptstyle
        \blacklozenge})$, $1 \times 10^{13}\, ({\scriptscriptstyle
        \square})$, $4 \times 10^{13}\, ({\scriptscriptstyle
        \blacksquare})$, $2 \times 10^{14}\, (\circ)$, $6 \times
      10^{14}\, (\bullet)$, $3 \times 10^{15}\, ({\scriptstyle
        \triangle})$, and $1 \times 10^{16}\, (\blacktriangle)$.}
    \label{fig:AfscaledPEC}
  \end{center}
\end{figure*}

While the predictions of PEC regarding the scaling properties and the
asymptotic behavior of the scaling functions are fulfilled,
$\Pi(t+t_w,t_w)$ and its PEC equivalent $\tilde \Pi(t+t_w,t_w)$ are
different in quantitative terms.  To see this, we have computed
$\tilde \Pi(t+t_w,t_w)$ numerically according to eq.~(\ref{eq:pitilde})
and compared it with $\Pi(t+t_w,t_w)$ obtained from the Monte--Carlo
simulations.  As shown in fig.~\ref{fig:AfscaledPEC} the
scaling functions associated with $\Pi(t+t_w,t_w)$ and $\tilde
\Pi(t+t_w,t_w)$ differ by a factor in the scaling variable and the
precise form in a transient region between the asymptotic regimes for
small and large arguments.

In summary we can conclude that all predictions of the PEC concerning the
scaling properties can be corroborated by the simulations. The PEC thus turns
out to be a powerful tool to uncover the mechanisms of aging in quenched
random energy landscapes.

\section{Exact evaluation of the Partial equilibrium formula}
\label{sec:exacteval}

In this part we show how the scaling arguments presented in
sec.~\ref{sec:scalingargument} can be validated by an exact evaluation
of the PEC formula (\ref{eq:pitilde}) in the limit of large $S$ (large
$t_w$). The reader who is not interested in these more mathematical
derivations, may skip this section and proceed with the summary in
sec.~\ref{sec:summary}.

When we replace the denominator of eq.~(\ref{eq:pitilde}) by
$\int_0^\infty d\lambda \exp(-\lambda \sum_{k=1}^S \tau_k)$ and notice
that all but three random $\tau_k$ are uncorrelated with the
$\tau_{j}$, $\tau_{n_j}$ appearing in the numerator of
(\ref{eq:pitilde}), we find
\begin{equation}
  \label{eq:pitS}
  \tilde \Pi(t,S) = S \int_0^\infty d\lambda\, \left\langle e^{-\lambda\,
      \tau}\right\rangle^{S-3}\, g(t; \lambda)\ , 
\end{equation}
where $g(t;\lambda)$ is defined by
\begin{eqnarray}
  \label{eq:gtlambda}
  g(t; \lambda) &\equiv& \left\langle \tau\, \exp\left( -t\, \tau^{\alpha-1}
      \sum_{j=1}^{2d} \tau_j^\alpha -\lambda (\tau + \tau_1 + \tau_2) \right)
      \right\rangle \nonumber \\ 
      &=& \int_1^\infty \frac{\theta\, d\tau}{\tau^\theta}\, e^{-\lambda\,
      \tau}\!\! \int_1^\infty \frac{\theta\, d\tau_1}{\tau_1^{1+\theta}}\,
      e^{-\lambda\, \tau_1}\!\! \int_1^\infty \frac{\theta\,
      d\tau_2}{\tau_2^{1+\theta}}\,  e^{-\lambda\, \tau_2} \nonumber
      \\ 
    && {} \times \exp\left( -t \frac{\tau_1^\alpha +
      \tau_2^{\alpha}}{\tau^{1-\alpha}} \right) \left[ f\left(
      \frac{t}{\tau^{1-\alpha}}\right)\right]^{2(d-1)} 
\end{eqnarray}
and
\begin{subequations}
  \label{eq:fz}
  \begin{equation}
    f(z) \equiv \int_1^\infty \frac{\theta\, d\tau}{\tau^{1+\theta}}\, e^{- z\,
      \tau^{\alpha}} = \frac{\theta}{\alpha}\, z^{\theta/\alpha}\, \Gamma\left(
      - \frac{\theta}{\alpha}, z \right)\ .
  \end{equation}
Here $\Gamma(a,z)=\int_z^\infty dt\, e^{-t}t^{a-1}$
denotes the incomplete Gamma function.
Note that $0\le f(z)\le1$ and that for $z \to 0$
  \begin{equation}
    \label{eq:fzexpand}
    f(z) = 1 - \mathop{\Gamma}\left(1-\frac{\theta}{\alpha} \right)
    z^{\theta/\alpha} - \frac{\theta}{\theta-\alpha}\; z +
    \Landau\big(z^2\big)\ ,
  \end{equation}
where $\Gamma(a)=\Gamma(a,0)$ is the Gamma function.
For $z\to\infty$ ($|{\rm arg} z|<3\pi/2$)
  \begin{equation}
    \label{eq:fzexpandinf}
   f(z)\sim \frac{\theta}{\alpha}\frac{e^{-z}}{z}\ .
  \end{equation}
\end{subequations}

In the limit $S\to\infty$ ($t_w\to\infty$), the asymptotic form of
eq.~(\ref{eq:pitS}) is [see eq.~(\ref{eq:explambdatauS2})]
\begin{equation}
  \label{eq:piSasympt}
  \tilde \Pi(t, S) \sim S
  \int_0^\infty d\lambda\, e^{-\lambda^\theta\tilde S}\, 
g\big( t; \lambda \big)\ , 
\end{equation}
where 
\begin{equation}
\tilde S \equiv \kappa\, S\ ,\quad\kappa\equiv\Gamma(1-\theta)\ .
\label{eq:kappa}
\end{equation}

\subsection{First scaling regime: $F_1(\Lambda_1)$}

By substituting $u=\tilde S^{1/\theta}\lambda$ in
eq.~(\ref{eq:piSasympt}) and $v \equiv \tau/\tilde S^{1/\theta}$ in
eq.~(\ref{eq:gtlambda}) we obtain
\begin{align}
  \tilde \Pi(&t,\, S) \sim \frac{\theta}{\kappa} \int_0^\infty
  \!\!\! du\, e^{-u^\theta} \int_{\tilde S^{-1/\theta}}^\infty
  \frac{dv}{v^\theta}\, e^{-u\, v} \nonumber \\
  & {} \times \int_1^\infty \frac{\theta\,d\tau_1}{\tau_1^{1+\theta}}
  e^{-u \tau_1/ \tilde S^{1/\theta}} \int_1^\infty
  \frac{\theta\,d\tau_2}{\tau_2^{1+\theta}} e^{-u \tau_2/ \tilde S^{1/\theta}}
  \label{eq:startauswpre} \\
  & {} \times \exp\left( - t\,\tilde S^{-\frac{1-\alpha}{\theta}}\, 
    \frac{\tau_1^\alpha +
  \tau_2^\alpha}{v^{1-\alpha}}\right) \left[ f\left(
  \frac{t\,\tilde S^{-\frac{1-\alpha}{\theta}}}
{v^{1-\alpha}}\right)\right]^{2(d-1)}
  \!\!\!\!\!\!\!\!\!\!\!\!\! . \nonumber 
\end{align}
When keeping 
\begin{equation}
\Lambda_1\equiv t\,\tilde S^{-(1-\alpha)/\theta}
\end{equation}
fixed, (\ref{eq:startauswpre}) has a well defined limit for $\tilde
S\to\infty$.  Before taking this limit, however, let us note at this point
that if we would have considered all $\tau_{n_j}$ in eq.~(\ref{eq:pitilde}) to
belong to the Brownian path, we had obtained
\begin{align}
  \tilde \Pi(&t,\,S) \sim \frac{\theta}{\kappa} \int_0^\infty
  \!\!\! du\, e^{-u^\theta} \int_{\tilde S^{-1/\theta}}^\infty
  \frac{dv}{v^\theta}\, e^{-u\, v} \times \nonumber \\
  & {} \times \prod_{i=1}^{2d} \left[ \int_1^\infty
  \frac{\theta\,d\tau_i}{\tau_i^{1+\theta}}
  \exp\left( - \frac{u\, \tau_i}{\tilde S^{1/\theta}} - \Lambda_1\,
  \frac{\tau_i^\alpha}{v^{1-\alpha}}\right) \right]
\label{eq:startausw} 
\end{align}
instead of (\ref{eq:startauswpre}). Now,
$v^{-\theta}\exp(-u^\theta-uv)\prod_{j=1}^{2d}\theta\tau_j^{-1-\theta}$
is an integrable majorant for the integrand in both
eqs.~(\ref{eq:startauswpre},\ref{eq:startausw}).  Hence, by
\textsc{Lebesgue}'s theorem we obtain from both equations the same
scaling function
\begin{equation}
  \label{eq:F1}
  F_1(\Lambda_1) \equiv \frac{\theta}{\kappa} \int_0^\infty
  \!\!\!\!\! du\, e^{-u^\theta} \int_0^\infty \!
  \frac{dv}{v^\theta}\, e^{-u\, v} \left[ f\left(
  \frac{\Lambda_1}{v^{1-\alpha}}\right)\right]^{2d}\!\!\!\!\! .
\end{equation}
We can conclude that it makes no difference here whether we consider the path
of distinct visited sites to have a one--dimensional or a compact topology or
anything in between. As required by normalization, $F_1(0)=1$.

\subsection{Limit $\Lambda_1 \to \infty$}

When transforming to variables $\xi \equiv \Lambda_1^{1/(1-\alpha)} u$ and
$\zeta \equiv \lambda_1^{-1/(1-\alpha)} v$ in eq.~(\ref{eq:F1}), we find
\begin{eqnarray*}
  F_1(\Lambda_1) &=& \Lambda_1^{-\theta/(1-\alpha)}\, \frac{\theta}{\kappa}
  \int_0^\infty \!\!\! d\xi\, e^{-\Lambda_1^{-\theta/(1-\alpha)} \xi^\theta}
  \times \\
  && {} \times \int_0^\infty \!\!\! d\zeta\, \zeta^{-\theta}\, e^{-\xi\,
  \zeta}\, f^{2d}\big(\zeta^{-(1-\alpha)} \big)\ .
\end{eqnarray*}
Since the integrand in the limit $\Lambda_1\to\infty$
is an integrable majorant for all $\Lambda_1$, we can take the
$\Lambda_1\to\infty$ limit under the integral to obtain
\begin{subequations}
  \label{eq:F1inf}
\begin{equation}
    F_1(\Lambda_1) \sim c_\infty\, \Lambda_1^{-\theta/(1-\alpha)}\quad
    \text{for}\quad  \Lambda_1 \to \infty \ , 
\end{equation}
where
\begin{equation}
   c_\infty \equiv \frac{\theta}{\kappa}\,
      \int_0^\infty \!\!\! \frac{d\zeta}{\zeta^{1+\theta}}\,
      f^{2d}(\zeta^{-(1-\alpha)})\ .
\end{equation}
\end{subequations}
Note that this integral is well defined because of the asymptotic
behavior of $f(.)$ given in eqs.~(\ref{eq:fz}b,c).

\subsection{Limit $\Lambda_1 \to 0$}

Since $F_1(\Lambda_1)\to1$ for $\Lambda_1\to0$, it is convenient to consider
\begin{eqnarray}
1-F_1(\Lambda_1)&=&\frac{\theta}{\kappa}\,\int_0^\infty\!\!\! du\,
 e^{-u^\theta}\!\!\int_0^\infty\!\! \frac{dv}{v^\theta}\,e^{-uv}\nonumber\\
&&\hspace{3em}{}\times\Biggl[1-
f^{2d}\left(\frac{\Lambda_1}{v^{1-\alpha}}\right)\Biggr].
\label{eq:oneminusf1}
\end{eqnarray}
For $\theta<\alpha$ it follows from eqs.~(\ref{eq:fz}a-c)
$1-f(x)^{2d}=2d\Gamma(1-\theta/\alpha)x^{\theta/\alpha}\varphi_<(x)$, where
$\varphi_<(x)$ is a bounded function, $\varphi_<(x)<M_<$ 
for $0\le x<\infty$, with
$\varphi_<(x)\to1$ for $x\to0$.  Hence,
\begin{eqnarray}
1-F_1(\Lambda_1)&=&2d\,\Gamma\left(1-\frac{\theta}{\alpha}\right)\,
\frac{\theta}{\kappa}\,\Lambda^{\theta/\alpha}
\int_0^\infty\!\!\! du\, e^{-u^\theta}\nonumber\\
&&\hspace{2em}{}\times
\!\!\int_0^\infty\!\!\frac{dv}{v^{\theta/\alpha}}\,e^{-uv}
\varphi_<\left(\frac{\Lambda_1}{v^{1-\alpha}}\right)\ .
\label{eq:oneminusf1<}
\end{eqnarray}
Since $M_<v^{-\theta/\alpha}\exp(-uv-u^{\theta})$
is an integrable majorant of the integrand, we find
\begin{subequations}
\begin{equation}
1-F_1(\Lambda_1)\sim c_<\,\Lambda_1^{\theta/\alpha}\ , \quad
\Lambda_1 \to 0,\ \ \theta < \alpha \ ,
\end{equation}
where
\begin{eqnarray}
c_<&\equiv&\frac{2d\,\theta\,\Gamma\bigl(1-\frac{\theta}{\alpha}\bigr)}
{\Gamma(1-\theta)}
\int_0^\infty\!\!\! du\, e^{-u^\theta}\!\!
\int_0^\infty\!\! \frac{dv}{v^{\theta/\alpha}}\,
e^{-uv}\nonumber\\
&=&\frac{2d\,\Gamma\bigl(1-\frac{\theta}{\alpha}\bigr)^2
\Gamma\bigl(\frac{1}{\alpha}\bigr)}{\Gamma(1-\theta)\,}\ .
\label{eq:csmaller}
\end{eqnarray}
\end{subequations}

For $\theta>\alpha$ the $v$ integral in eq.~(\ref{eq:oneminusf1<}) would
become divergent when taking the $\Lambda_1\to0$ limit in
$\varphi_<(\Lambda_1/v^{(1-\alpha)})$.  We thus transform the $v$ variable in
eq.~(\ref{eq:oneminusf1}), $w\equiv \Lambda_1/ v^{1-\alpha}$,
to obtain
\begin{eqnarray}
1-F_1(\Lambda_1)&=&
\frac{\theta\,\Lambda_1^{\frac{1-\theta}{1-\alpha}}}{(1-\alpha)\,\kappa}\,
\int_0^\infty\!\!\! du\,e^{-u^\theta}\!\!
\int_0^\infty\!\! \frac{dw}{w^{1 +
      \frac{1-\theta}{1-\alpha}}}\,\nonumber\\
&&\times\bigl[1-f^{2d}(w)\bigr]\,
\exp\left(-\frac{u\,\Lambda_1^{\frac{1}{1-\alpha}}}{w^{\frac{1}{1-\alpha}}
             }\right)\,.
\label{eq:oneminusf1>}
\end{eqnarray}
From eqs.~(\ref{eq:fz}a-c) it follows for $\theta>\alpha$ that
$1-f^{2d}(w)=w\varphi_>(w)$ , where $\varphi_>(w)$
is a bounded function, $\varphi_>(w)<M_>$ for $0\le w<\infty$, with
$\varphi_>(w)\to(\theta-\alpha)/2d\theta$ for $w\to0$.  Hence,
$M_>\exp(-u^\theta)w^{(1-\theta)/(1-\alpha)}$ is an integrable majorant of the
integrand in (\ref{eq:oneminusf1>}) and we can take the $\Lambda_1\to0$
limit under the integral, yielding
\begin{subequations}
\begin{equation}
1-F_1(\Lambda_1)\sim c_>\,\Lambda_1^{\frac{1-\theta}{1-\alpha}}\quad
    \Lambda_1 \to 0,\ \ \theta > \alpha \ ,
\end{equation}
where
\begin{equation}
c_>=\frac{\Gamma\bigl(\frac{1}{\theta}\bigr)}
{(1-\alpha)\,\Gamma(1-\theta)}
\int_0^\infty\!\!\! \frac{dw}{w^{1 +\frac{1-\theta}{1-\alpha}}}\,
\bigl[1-f^{2d}(w)\bigr]\,.
\end{equation}
\end{subequations}

\subsection{Second scaling regime: $F_2(\Lambda_2)$}

The second scaling regime is more difficult to extract from
eqs.~(\ref{eq:pitS},\ref{eq:gtlambda}). We start by taking advantage
of the normalization $\tilde \Pi(t=0,S)\sim1$ in order to write
\begin{align}
  1-\tilde \Pi(t,\, S) \sim & \frac{\theta}{\kappa} \int_0^\infty
  \!\!\! du\, e^{-u^\theta} \int_{\tilde S^{-1/\theta}}^\infty
  \frac{dv}{v^\theta}\, e^{-u\, v} \label{eq:startauswagain} \\
  & \hspace*{-4em} \times
  \left(\prod_{j=1}^{2d}\int_1^\infty
\frac{\theta\,d\tau_j}{\tau_j^{1+\theta}}\right)\,
    \exp\left(-\frac{u}{\tilde S^{1/\theta}}\sum_{k=1}^{n}\tau_k\right)
  \nonumber \\
  & {} \times \left[1- \exp\left(-\frac{\Lambda_1}{v^{1-\alpha}}
      \sum_{j=1}^{2d}\tau_j^\alpha\right) \right]\ . \nonumber
\end{align}
This corresponds to eqs.~(\ref{eq:startauswpre},\ref{eq:startausw}),
if the number of neighboring sites belonging to the Brownian path is
$n$, $2\le n\le2d$ (see the discussion in sec.~\ref{sec:pec}).

After the transformation $v\to w=\Lambda_1 \sum_{j=1}^{2d}\tau_j^\alpha/
v^{(1-\alpha)}$ this can be rewritten as
\begin{align*}
  1-&\tilde \Pi(t,\, S) \sim \frac{\theta\Lambda_1}{(1-\alpha)\kappa}
  \int_0^\infty
  \!\!\! du\, e^{-u^\theta} u^{-(\alpha-\theta)}\\
  & {} \times
  \left(\prod_{j=1}^{2d}\int_1^\infty
\frac{\theta\,d\tau_j}{\tau_j^{1+\theta}}\right)\,
    \exp\left(-\frac{u}{\tilde S^{1/\theta}}\sum_{k=1}^{n}\tau_k\right)
  \left(\sum_{j=1}^{2d}\tau_j^\alpha\right)\\
  & {} \times \left[\Lambda_1\,u^{1-\alpha}\sum_{j=1}^{2d}\tau_j^\alpha 
    \right]^{\frac{1-\theta}{1-\alpha}-1}
  \!\!\!\int_0^{t\sum_{j=1}^{2d}\tau_j^\alpha}\!\!
  \frac{dw\,(1-e^{-w})}{w^{1+\frac{1-\theta}{1-\alpha}}}\\
  & {} \times \exp\left[-\left(\frac{\Lambda_1\,u^{1-\alpha}
        \sum_{j=1}^{2d}\tau_j^\alpha
        }{w}\right)^{\frac{1}{(1-\alpha)}}\right].
\end{align*}
We now decompose this expression into the sum of two parts
corresponding to a decomposition $\sum_{j=1}^{2d}\tau_j^\alpha=
\sum_{j=1}^{n}\tau_j^\alpha+\sum_{j=n+1}^{2d}\tau_j^\alpha$ in the
second line. We can use the symmetry with respect to $\{\tau_j\}$ in
order to replace $\sum_{j=1}^n\tau_j^\alpha$ by $n\tau_1^\alpha$ in the
first part and $\sum_{j=n+1}^{2d}\tau_j^\alpha$
by $(2d-n)\tau_{2d}^\alpha$ in the second part. After
the transformation $\zeta=\tilde S^{-1/\theta}\tau_1$ in the
first part, and the transformation $\zeta=\tilde
S^{-1/\theta}\tau_{2d}$ in the second part one can, for
$\theta<\alpha$, again use \textsc{Lebesgue}'s theorem to perform the
limit $S\to\infty$ for fixed
\begin{equation}
\Lambda_2\equiv
\tilde S^{\alpha/\theta}\Lambda_1=\frac{t}{\tilde S^{(1-2\alpha)/\theta}}\,,
\quad\theta<\alpha<\frac{1}{2}\,.\label{eq:lambda2here}
\end{equation}
This yields
\begin{subequations}
\begin{align}
&\tilde S\,[1-\Pi(t,S)]\sim F_2(\Lambda_2)\\[0.5em]
&F_2(\Lambda_2)=n\, F_2^{(1)}(\Lambda_2)+
(2d-n)\,F_2^{(2)}(\Lambda_2)\,,
\label{eq:f2sum}
\end{align}
\end{subequations}
where
\begin{align}
F_2^{(1)}(\Lambda_2)&\equiv\frac{\theta^2\,\Lambda_2}{(1-\alpha)\,\kappa}
\int_0^\infty \!\!\frac{du\, e^{-u^\theta}}{u^{\alpha-\theta}}\, 
\nonumber\\[0.5em]
&\hspace{1em}\times\int_0^\infty\!\!\!
\frac{d\zeta e^{-u\zeta}}{\zeta^{1-(\alpha-\theta)}}\,
h(\Lambda_2\,u^{1-\alpha}\,\zeta^\alpha)\,,
\label{eq:f21here}\\[0.5em]
&\hspace{-3em}h(x)\equiv x^{\frac{1-\theta}{1-\alpha}-1}\int_0^\infty
\frac{dw\,(1-e^{-w})}{w^{1+\frac{1-\theta}{1-\alpha}}}\,
e^{-(x/w)^{1/(1-\alpha)}}\hspace{-0.5em},
\end{align}
and
\begin{align}
F_2^{(2)}(\Lambda_2)&\equiv\frac{\theta^2\,\Lambda_2}{(1-\alpha)\,\kappa}
\int_0^\infty \!\!\frac{du\, e^{-u^\theta}}{u^{\alpha-\theta}}
\!\!\int_0^\infty\!
\frac{d\zeta\, h(\Lambda_2\,u^{1-\alpha}\,\zeta^\alpha)}
{\zeta^{1-(\alpha-\theta)}}
\nonumber\\[0.5em]
&=\frac{\theta^2\,\Lambda_2^{\theta/\alpha}}
  {\alpha(1-\alpha)\,\kappa} \int_0^\infty\!\!
  \frac{du\, e^{-u^\theta}}{u^{1-\theta/\alpha}}\,
  \!\!\int_0^\infty\!\!\nonumber\frac{dv\, h(v)}
  {v^{\theta/\alpha}}\\[0.5em]
&=\frac{\Gamma\bigl(1-\frac{\theta}{\alpha}\bigr)^2
\Gamma\bigl(\frac{1}{\alpha}\bigr)}{\Gamma(1-\theta)\,}\,
\Lambda_2^{\theta/\alpha}\,.\label{eq:f22here}
\end{align}
The restriction to $\alpha<1/2$ in (\ref{eq:lambda2here})
follows from the fact that in order for the scaling regime to be
relevant, $t_2\sim\tilde S^{(1-2\alpha)/\theta}\sim
t_w^{\gamma(1-2\alpha)/\theta}$ should increase with increasing $t_w$.
The function $h(x)$ has the asymptotic behavior
\begin{equation}
h(x)\sim\left\{\begin{array}{l@{\hspace{1em}}l}
(1-\alpha)\,\Gamma(\alpha-\theta)\ , & x\to0^+\\[1.5ex]
(1-\alpha)\,\Gamma(1-\theta)\,x^{-1}\ , & x\to\infty\end{array}\right.\ .
\end{equation}
The importance of the condition $\theta<\alpha$ now becomes clear, since for
the $\zeta$ integrand to be integrable in
eqs.~(\ref{eq:f21here},\ref{eq:f22here}) for $\zeta\to0$, $\alpha-\theta$ has
to be positive.

\subsection{Limit $\Lambda_2\to\infty$}
After the transformation $\zeta\to v=\Lambda_2
u^{(1-\alpha)}\zeta^\alpha$, eq.~(\ref{eq:f21here}) gives
\begin{align}
  F_2^{(1)}(\Lambda_2)&=\frac{\theta^2\,\Lambda_2^{\theta/\alpha}}
  {\alpha(1-\alpha)\,\kappa} \int_0^\infty
  \frac{du}{u^{1-\theta/\alpha}}\,
  e^{-u^\theta}\nonumber\\
  &{}\times \int_0^\infty \frac{dv}
  {v^{\theta/\alpha}}e^{-u\,v^{1/\alpha}\,
\Lambda_2^{-1/\alpha}u^{-(1-\alpha)/\alpha}}\,h(v)\,.
\end{align}
The limit $\Lambda_2\to\infty$ can then be taken under the
integral, yielding [cf.\ eq.~(\ref{eq:f22here})]
\begin{equation}
F_2^{(1)}(\Lambda_2)\sim F_2^{(2)}(\Lambda_2)\,.
\end{equation}
Using eqs.~(\ref{eq:f2sum},\ref{eq:f22here}) we finally obtain
\begin{equation}
F_2(\Lambda_2)\sim c_2^{(\infty)}\Lambda_2^{\theta/\alpha}\,,
\quad
c_2^{(\infty)}\equiv \frac{2d\,\Gamma\bigl(1-\frac{\theta}{\alpha}\bigr)^2
\Gamma\bigl(\frac{1}{\alpha}\bigr)}{\Gamma(1-\theta)\,}\,.
\end{equation}
As required by matching, $c_2^{(\infty)}=c_<$ [cf.\ eq.~(\ref{eq:csmaller})].

\subsection{Limit $\Lambda_2\to0$}
\label{subsec:l2limit}
Since $h(x)$ is monotonously decreasing with $x$ for $x>0$,
we can replace $h(\Lambda_2u^{1-\alpha}\zeta^{\alpha})$
by $h(0)=(1-\alpha)\Gamma(\alpha-\theta)$
in eq.~(\ref{eq:f21here}). Hence,
\begin{equation}
  F_2^{(1)}(\Lambda_2)\sim c_2 \Lambda_2\ ,
  \label{eq:f21lambdasmall}
\end{equation}
where
\begin{eqnarray}
c_2 &\equiv&
\frac{\theta^2\,\Gamma(\alpha-\theta)\,}
{\kappa} \int_0^\infty\!\! \frac{du\, e^{-u^\theta}}{u^{\alpha-\theta}}\,
\!\!\int_0^\infty\!\! \frac{d\zeta\, e^{-u\zeta}}
{\zeta^{1-(\alpha-\theta)}}\nonumber\\
&=&\frac{\theta}{1-\alpha}\frac{\Gamma(\alpha-\theta)
\Gamma\left(\frac{1-2(\alpha-\theta)}{\theta}\right)}{\Gamma(1-\theta)}\ .
\end{eqnarray}
This means that for small $\Lambda_2$, $F_2(\Lambda_2)$
should be dominated by $F_2^{(2)}(\Lambda_2)$, except
for $d>1$ (where $F_2(\Lambda_2)=2F_2^{(1)}(\Lambda_2)$).
For a given (mean) value of $n$ we thus find
\begin{equation}
F_2(\Lambda_2)\sim c_2^{(0)}\Lambda_2^{\theta/\alpha}\ ,\quad
c_2^{(0)}\equiv \left(1-\frac{n}{2d}\right) c_2^{(\infty)}\,. 
\end{equation}
This $\Lambda_2\to0$ limit of the generalized scaling form depends on the
number of neighbors being considered to belong to the Brownian path (see the
discussion in sec.~\ref{sec:pec}). According to eq.~(\ref{eq:f21lambdasmall}),
there can occur an intermediate regime, where $F_2(\Lambda_2)$ depends
linearly on $\Lambda_2$. This intermediate regime seems to be more pronounced
for the ``true dynamics'' (see Fig.~\ref{fig:Afscaled2}) than for the dynamics
predicted by the PEC formula (\ref{eq:pitilde}).

\section{Summary and Conclusions}
\label{sec:summary}

We have studied aging within the framework of a simple hopping model
mimicking a system that performs thermally activated transitions
between the deep free energy minima of its configuration space.  Based
on general arguments from the statistical theory of extremes we have
chosen the free energy density of states to exhibit an exponential
tail.  In order to effectively quantify the influence of the initial
and target site on the energy barrier to be surmounted during a
transition, we introduced a parameter $\alpha$, $0\le\alpha<1$, in the
hopping rates that turned out to strongly influence the aging
properties. We have found that generically, subaging occurs in these
models, an effect related to the multiple visits of deep traps. We
have also found that different time scales, corresponding to different
scaling regions, appear in these models.

These aging properties can be understood from a partial equilibrium concept
(PEC) that, despite not being an exact quantitative description, provides a
powerful tool to study the scaling properties of the aging dynamics. Based on
the PEC we first motivated the occurrence of subaging behavior and generalized
scaling forms in terms of simple scaling arguments. We then presented a
detailed analysis of the PEC formula (\ref{eq:pitilde}) and calculated the
aging functions following from (\ref{eq:pitilde}) and their asymptotics
exactly. With respect to the scaling properties the predictions could be
confirmed by Monte-Carlo simulations in $d=1, 10, 100$ and 1000 dimensions.

The fact that even for $d=1000$ the ``quenched model'' has aging
properties different from the ``annealed model'' studied earlier in
\cite{Monthus/Bouchaud:1996} is rather surprising, since the number of
distinct visited sites $S$ in $d>2$ scales as the number of all
transitions $N$ between minima for large $N$, $S\sim N$. From this one
tends to conclude that the system effectively explores a new minimum
in each transition, which would correspond to the annealed situation.
However, in the quenched situation one can imagine that there is
always some {\it local} equilibrium established at the site with
minimal energy reached after time $t_w$, and this local equilibration
effect slows down the diffusion in configuration space on all time
scales, i.e.\ instead of $S\sim N\sim t_w^{\theta'}$ with
$\theta'=\min[1,\theta/(1-\alpha)]$ in the annealed situation we have
$S\sim N\sim t_w^{\theta}$ in the quenched situation for $d>2$,
$0<\theta<1$.

The existence of a local equilibrium around the ``dominant'' site with
minimal energy after time $t_w$ does not imply that there must be a
true equilibration on all visited sites as it is assumed in the PEC
formula (\ref{eq:pitilde}). In fact, by studying the disorder averaged
participation ratios $Y_q(t)\equiv\langle\sum_j P_j(t)^q\rangle$,
where $P_j(t)$ is the probability for the system to be at minimum $j$
at time $t$, we find that the PEC never becomes exact in the limit
$t\to\infty$, not even in $d=1$ where each trap is visited an infinite
number of times. This behavior offers the possibility to define an
effective temperature in the non--equilibrium aging regime, which
enters a modified fluctuation--dissipation theorem, similar as it was
found for mean--field spin glass models
\cite{Cugliandolo/Kurchan:1994}. The role of an effective temperature
in the landscape model considered here will be discussed elsewhere
\cite{Bertin/etal:2001} (for recent progress in our understanding of
this problem coming from MD simulations, see
\cite{Sciortino/Tartaglia:2001}).

With respect to the applicability of the analysis outlined above the
question arises, whether the characteristics of the aging dynamics can
be worked out also for general hopping rates not exhibiting the
specific form given in eq.~(\ref{eq:wij}). For the PEC to be
applicable, the system should have the tendency to approach
equilibrium (that truly exists only for $\theta>1$), so that one may
require the jump rates to obey detailed balance. It is then indeed
straightforward, by using the simple arguments presented in
sec.~\ref{sec:scalingargument}, to extract all characteristic time
scales $t_j(t_w)$ and to predict the scaling properties. A necessary
ingredient for this procedure to work correctly, however, is the
robustness of the scaling relation $S(t_w)\sim t_w^\gamma$ [cf.\ 
eqs.~(\ref{eq:stwgamma},b)]. Preliminary results indicate that for
``physical choices'' of the jump rates (meaning that the dependence of
the energies of the initial and target site on the saddle point energy
is reasonable), eqs.~(\ref{eq:stwgamma},b) always hold true.

Moreover, it is possible also to consider some random distribution of
the {\it form} of the jump rates (as it is expected to occur when the
dynamics in configuration space is mapped onto a jump process by means
of some quantitative analysis), and to work out the aging features of
such more realistic models. It turns out then, that in principle
infinitely many aging regimes can exists in the two-time plane
$t,t_w\ge0$.  A thorough discussion of these issues, however, is
beyond the scope of the present work.

\begin{acknowledgments}
  We should like to thank W.~Dieterich and E.~Bertin for stimulating
  discussions. P.M. gratefully acknowledges financial support from the
  Deutsche Forschungsgemeinschaft by a Heisenberg fellowship
  (Ma~1636/2-1).
\end{acknowledgments}

\begin{appendix}

\section{Technical details of the Monte--Carlo simulations}
\label{sec:mc}

We use the standard continuous--time Monte Carlo algorithm as
discussed in detail e.g.\ in \cite{Binder/Heermann:1992} to simulate
the stochastic process defined in sec.~\ref{sec:model}.  A special
problem arises for large dimensions $d\gg1$, where it is not possible
to save the energies within a hypercube of even small linear
dimension.  To resolve this problem, we use hash--maps, as, for
example, the \verb#hash_map# template provided by the Standard
Template Library of ISO--C++.

The hash--function should be computable quickly
and at the same time the sites $\boldsymbol{x}$ being encountered must be
mapped to different hash--values as often as possible. 
For dimensions $d\ge10$ we found
\begin{equation}
  \label{eq:hashfunc}
  f(\boldsymbol{x}) = \sum_{n=1}^{2d} n\, x_n
\end{equation}
to do a good job. For large $d$ the RAM consumption of the computer
programs is the limiting factor when trying to access longer times
$t_w$. While simulations for $d\le 10$ can easily be performed on
workstations, for $d \ge 100$ computers with 4GB of RAM and more are
necessary.

\section{Connection between the ``number of distinct visited sites''
  and the waiting time}\label{sec:properties}

For $\alpha=0$, i.e.\ the trap model, we use a scaling argument
discussed in \cite{Harder/etal:1987} to derive the behavior of
$S(t_w)$ for large $t_w$.  Then we show by a finite size scaling
argument that in $d=1$ the behavior of $S(t_w)$ for $\alpha=0$ is not
expected to change for $0<\alpha<1$. Furthermore, we give general
arguments for the invariance of eq.~(\ref{eq:gamma}) with
respect to $\alpha$ for all $d$. Finally we confirm eqs.~(\ref{eq:stwgamma},b)
by Monte-Carlo simulations.

\subsection{Trap model ($\alpha = 0$)}
\label{subsec:alpha0}

After $N\gg1$ transitions of the system, the typical elapsed time $t_w$ is
\begin{align}
  \label{eq:tNrel}
  t_w(N) &\simeq \sum_{i=1}^{N} \tau_{i} \simeq N \sum_{\mu} \frac{g_\mu(N)}{N}
  \tau_\mu\Delta\tau_\mu\nonumber\\
&\hspace{-1em}{}\sim N \!\!\!\!\!\!\!\!
  \int\limits_{1}^{\taumax(S(N))} \!\!\!\!\!\!\!\! d\tau\,
  \tau\, \rho(\tau)\sim N\, [\taumax(S(N))]^{(1-\theta)}\ ,
\end{align}
where $g_\mu(N)$ is the typical number of $\tau_j$ falling in some
interval
$\tau_\mu-\Delta\tau_\mu/2\le\tau_\mu\le\tau_\mu+\Delta\tau_\mu/2$, 
$S(N)$ is the typical number of distinct visited sites after $N$ jumps
and $\taumax(S)$ is the typical
maximal $\tau$ obtained after
the system encountered $S$ distinct sites. 

Since $\taumax(S) \sim S^{1/\theta}$
and (see e.\,g.\ \cite{Hughes:1995})
\begin{equation}
  \label{eq:SNrel}
  S(N) \sim \left\{ \begin{array}{c@{\hspace{2em}}c} 
      N^{d/2}\ , & 1\le d < 2 \\[0.2em]
      N/\ln N\ , & d = 2 \\[0.2em]
      N\,  & d > 2 
    \end{array} \right.\ ,
\end{equation}
we find
\begin{equation}
  \label{eq:tSrel}
  t_w(S) \sim \left\{ \begin{array}{c@{\hspace{2em}}c} 
      S^{[d+(2-d)\theta]/d\theta}\ , & 1\le d < 2 \\[0.2em]
      S^{1/\theta}\ln S & d = 2 \\[0.2em]
      S^{1/\theta}\ , & d > 2 
    \end{array} \right.\ .
\end{equation}
This yields eqs.~(\ref{eq:stwgamma},b) for $\alpha=0$. 

We note that in the annealed model eq.~(\ref{eq:SNrel}) remains valid,
while $\taumax\sim N^{1/\theta}$, leading to $S(t_w)\sim
t_w^{d\theta/2}$ for $1\le d<2$, $S(t_w)\sim t_w^\theta/\ln t_w$ for
$d=2$, and $S(t_w)\sim t_w^\theta$ for $d>2$. Due to our discussion in
sec.~\ref{sec:annealedcase}, one can replace $\theta$ by
$\theta'=\min(1,\theta/(1-\alpha))$ in these formulae to obtain the
behavior for $0\le\alpha<1$.

\subsection{One dimension ($0\le\alpha<1$)}

Let us consider a finite chain with $L$ sites
and site energies distributed according to eq.~(\ref{eq:Edistribution}).
Then the mean square
displacement 
$\langle \Delta x^2(t) \rangle$ of a particle performing a random
walk on this chain with the hopping rates given in (\ref{eq:wij}) 
is expected to scale as
\begin{equation}
\label{eq:x2t}
  \langle \Delta x^2(t) \rangle \sim \left\{ 
    \begin{array}{l@{\quad\text{for}\quad}l}
      t^\eta & \langle \Delta x^2(t) \rangle \ll L^2 \\[0.2em]
      D(L)\, t & \langle \Delta x^2(t) \rangle \gg L^2
    \end{array}
  \right.
\end{equation}
for large $L$. The diffusion coefficient $D(L)$ can be written as
\cite{Kutner/etal:1994}
\begin{equation}
  D(L) = \frac{L^2}{\sum_{j=1}^{L} \big(p_j^{\text{(eq)}}
  w_{j,j+1}\big)^{-1}}
\end{equation}
with $p_j^{\text{(eq)}}=\exp(-\beta E_j)/\sum_{k=1}^L\exp(-\beta E_k)=
\tau_j/\sum_k\tau_k$. The denominator then reads, using
(\ref{eq:wijtau}), $\sum_{i=1}^L \tau_i \sum_{j=1}^L
(\tau_j\tau_{j+1})^{-\alpha}$. Since $\langle ( \tau_j\, \tau_{j+1}
)^{-\alpha} \rangle$ exists (for $\alpha>-\theta$), the second sum
gives a contribution $\propto L$ for large $L$, while the first sum
has no finite average and is dominated by the maximum $\taumax \sim
L^{1/\theta}$. Thus we find
\begin{equation}
  \label{eq:DL2}
  D(L) \sim \frac{L^2}{L^{1/\theta}\, L} \sim L^{1-1/\theta}\ .
\end{equation}
At the crossover time $t_x$, where $\langle \Delta
x^2(t_x) \rangle$ changes its behavior in eq.~(\ref{eq:x2t}), we obtain from 
continuity
\begin{equation}
t_x^\eta\sim D(L)t_x\sim L^2\ .
\end{equation}
This implies $t_x\sim L^{2/\eta}$ and $t_x\sim L^{1+1/\theta}$
yielding
\begin{equation}
  \eta = \frac{2\, \theta}{1+\theta}\ .
\end{equation}
Since $S(t_w)\sim\langle \Delta x^2(t) \rangle^{1/2}$, we finally
obtain for $0\le\alpha<1$
\begin{equation}
  S(t_w) \sim t_w^{\eta/2} = t_w^{\theta/(1+\theta)}
\end{equation}
in agreement with eqs.~(\ref{eq:stwgamma},b).

\subsection{General arguments}
\label{sec:genarg}

The very physical difference between the trap model ($\alpha=0$) and
models with weighted rates ($\alpha>0$) is the occurrence of
forward--backward jump correlations.  When the system jumps from a
site with low energy to a site with energy close to zero (such
energies are most likely), it has high tendency to jump back for
$\alpha>0$. More generally, when the system enters a region of
connected low energy sites, it will, before escaping this region,
perform more and more jumps between the low energy sites the larger
the value of $\alpha$ is. Once it left the region, it again has high
tendency to jump back to it.

One may regard a cluster of sites with deep energies and the
surrounding shell of sites with higher energy as a ``supertrap''. On a
coarse--grained level with respect to time, the particle performs
``superhops'' between these supertraps. Then the essential difference
between the $\alpha=0$ and the $\alpha>0$ situation disappears, since
there are no increased backward jump correlations between the
superhops. One thus suspects that $\alpha$ only rescales the time
$t_w$ of the $S(t_w)$ relation but does not change its exponent.
Indeed this is what we have shown more explicitly in $d=1$ in the
previous subsection, and there is no reason why the argument should
fail in higher dimensions.

It is worth to note that the same arguments also apply to
$\Pi(t+t_w,t_w)$, if one generalizes it to a quantity
$\Pi_n(t+t_w,t_w)$ which is defined as the averaged probability that
the system after a waiting time $t_w$ does not leave a region of
radius $n$ in configuration space.  Clearly, $\Pi(t+t_w,t_w) =
\Pi_0(t+t_w,t_w)$, but for larger $n$ only superhops should lead to a
decrease in $\Pi_n(t+t_w,t_w)$.  In fact we found that for $n \ge 1$
$\Pi_n(t+t_w,t_w)$ shows normal aging, i.e.\ $\Pi_n(t+t_w,t_w)\sim
F_1^{(n)}(t/t_w)$ for $\alpha>0$.

\subsection{Monte--Carlo results}
\label{subsec:ndvsnum}
\begin{figure}[htbp]
  \begin{center}
    \hspace*{-0.7cm}
    \epsfig{file=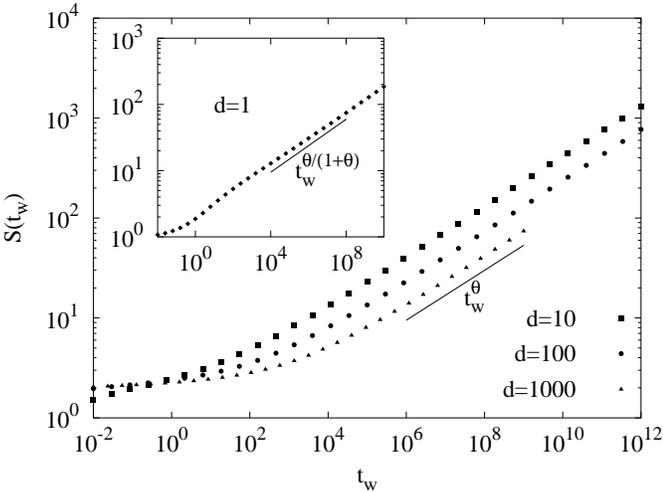,width=9.5cm}
    \caption{Number of distinct visited sites $S(t_w)$ for
      $(\theta,\alpha)=(1/4,3/8)$ and different dimensions $d$.}
    \label{fig:ndvsa}
  \end{center}
\end{figure}

\begin{figure}[htbp]
  \begin{center}
    \hspace*{-0.7cm}
    \epsfig{file=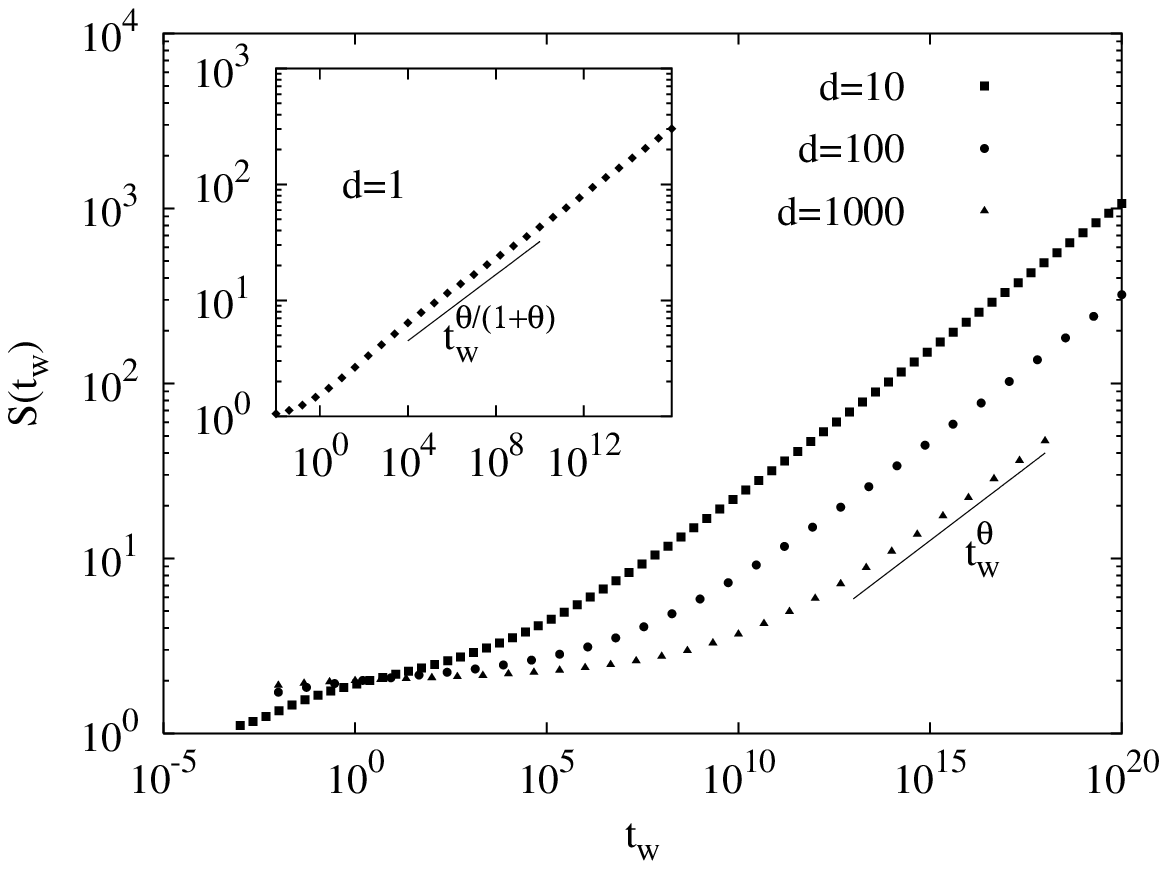,width=9.5cm}
    \caption{Number of distinct visited sites $S(t_w)$ for
      $(\theta,\alpha)=(1/6,1/4)$ and different dimensions $d$.}
    \label{fig:ndvsb}
  \end{center}
\end{figure}

Figs.~\ref{fig:ndvsa} and \ref{fig:ndvsb} show $S(t_w)$ from Monte
Carlo simulations in $d=1$, 10, 100, and 1000
for parameter sets $(\theta,\alpha) = (1/4,3/8)$ and
$(1/6,1/4)$, respectively. In all cases $S(t_w)$ shows the behavior
predicted by eqs.~(\ref{eq:stwgamma},b).

\section{$\tilde \Pi(t,S)$ in the limit $S \to \infty$}
\label{sec:explambdatau}

To derive the large $S$ limit of eq.~(\ref{eq:pitS}),
\begin{equation}
  \label{eq:explambdatauS}
  \tilde \Pi(t,S) = S \int_0^\infty d\lambda\, \left\langle
  e^{-\lambda\, \tau} \right\rangle^{S-3} g(t; \lambda)\ ,
\end{equation}
we need to consider the small $\lambda$ limit of
\begin{eqnarray*}
  \left\langle e^{-\lambda\, \tau}\right\rangle &=&
  \int_1^{\infty} \frac{\theta d\tau}{\tau^{1+\theta}}\, e^{-\lambda\,
    \tau}
  =\theta\lambda^\theta\Gamma(-\theta,\lambda)\\
  &=& 1 - \Gamma(1-\theta)\, \lambda^{\theta} 
  + \frac{\theta}{1-\theta}\, \lambda + \Landau(\lambda^2)\\
  &\equiv& e^{- \varphi(\lambda)\, \lambda^\theta }
\end{eqnarray*}
where $\varphi(\lambda)$ is a continuous function with
$\varphi(\lambda) \to \Gamma(1-\theta)$ for $\lambda \to 0$.
Furthermore, $\left\langle e^{-\lambda\, \tau}\right\rangle$ has an
upper bound $e^{- a\, \lambda^\theta}$ for $\lambda\ge0$,
\begin{equation}
  \label{eq:explambdatau-majorant}
  \left\langle e^{-\lambda\, \tau}\right\rangle < e^{- a\,
  \lambda^\theta}
\end{equation}
with some constant $a$, $0<a\le1$, being independent of $\lambda$. To
proof this for $\lambda \ge 1$ we compare
\begin{align}
  e^{- a\, \lambda^\theta} &= \int_1^\infty \!\!\! d\tau\, \left [
  a \lambda^\theta\, \tau^{-1+\theta}\, e^{- a (\lambda
    \tau)^\theta} \right] \nonumber
\intertext{with}
  \left\langle e^{-\lambda\, \tau}\right\rangle &= \int_1^\infty
  \!\!\! d\tau\, \left[ \theta\, \tau^{-1-\theta}\, e^{- \lambda \tau}
  \right]\nonumber\ .
\end{align}
For $\theta \le a \le 1$ and $\tau \ge 1$ the first integrand is
larger than the second integrand for all $\lambda \ge 1$, thus the
first integral is larger than the second integral. Since $e^{-a\,
  \lambda^\theta}$ for fixed $\lambda$ is strictly monotonously
decreasing with $a$ this remains valid also for $0 < a < \theta$. To
proof property (\ref{eq:explambdatau-majorant}) for $\lambda < 1$ we
note that for small argument $\epsilon$ it is
\begin{align}
  e^{- \varphi(\epsilon)\, \epsilon^\theta } + \Landau(\epsilon) &=
  1-\Gamma(1-\theta)\, \epsilon^\theta \nonumber\\
  &< 1 - a\, \epsilon^\theta
  = e^{- a\, \epsilon^\theta} +
  \Landau(\epsilon^{2\theta})\nonumber \ ,
\end{align}
since $\Gamma(1-\theta) > 1$ for $\theta < 1$. We can deduce that it
exists a finite interval $(0,\lambda_0]$ where
eq.~(\ref{eq:explambdatau-majorant}) holds. Because $\left\langle
  e^{-\lambda\, \tau}\right\rangle $ is strictly monotonously
decreasing it is $\left\langle e^{-\lambda_0\, \tau}\right\rangle <
1$. When choosing $a_0 \equiv - \ln \left\langle e^{-\lambda_0\,
    \tau}\right\rangle > 0$ it holds that
\[
  \left\langle e^{-\lambda\, \tau}\right\rangle < e^{-a_0\,
  \lambda^\theta}\qquad \text{for}\quad 0 < \lambda \le 1
\]
and the proof of eq.~(\ref{eq:explambdatau-majorant}) is complete.
 
With the transformation $\lambda \to u \equiv S^{1/\theta} \lambda$
eq.~(\ref{eq:explambdatauS}) gives
\begin{align}
  \tilde \Pi(t,S) = \int_0^\infty \!\!\! du\, &\exp\left(
    - \varphi\big(u^\theta / S \big)\, \frac{S-3}{S}\, u^\theta
  \right) \nonumber\\
  &  {} \times S^{1-1/\theta}\,
  g\!\left(t;\frac{u}{S^{1/\theta}}\right)\ . \nonumber
\end{align}
When using eq.~(\ref{eq:explambdatau-majorant}) we can estimate
\begin{eqnarray*}
  \exp\!\left( - \varphi\big(u^\theta / S \big) \, \frac{S-3}{S}\,
    u^\theta \right)
  &<& \exp\!\left( -a\, \frac{S-3}{S}\, u^\theta \right)\\
  &<& \exp\!\left( -\frac{a}{2}\, u^\theta \right)
\end{eqnarray*}
for $S > 6$ and from eq.~(\ref{eq:gtlambda}),
\begin{eqnarray*}
  S^{1-1/\theta}\, g(t;S^{-1/\theta}\, u) &\le&
  S^{1-1/\theta}\,  \int_1^\infty\frac{\theta d\tau}{\tau^{\theta}}
  e^{-S^{-1/\theta}\, u\, \tau}\\ 
  &=& S^{1-1/\theta} \theta \lambda^{\theta-1}
  \Gamma(1-\theta,S^{-1/\theta}\, u)\\
  &\le& \theta\Gamma(1-\theta) u^{\theta-1}\ .
\end{eqnarray*}
These estimations will allow us to use \textsc{Lebesgue}'s theorem
when considering the limit $S \to \infty$ in
eqs.~(\ref{eq:startauswpre},\ref{eq:startauswagain}). Thus we can
write asymptotically
\begin{align}
  \label{eq:explambdatauS2}
  \tilde \Pi(t,S) &\sim S^{1-1/\theta} \int_0^\infty \!\!\! du\,
    e^{-\kappa\, u^\theta}\, g\big(t;S^{-1/\theta}\, u \big)\nonumber\\
  &\sim S \int_0^\infty \!\!\! d\lambda\, e^{-\lambda^\theta
    \kappa\, S}\,  g\big( t; \lambda \big)\ ,
\end{align}
where the shortcut $\kappa=\Gamma(1-\theta)$ [cf.\ 
eq.~(\ref{eq:kappa})] has been used. 

\end{appendix}

\end{document}